\documentclass[twocolumn]{aastex63}

\usepackage{bm}
\usepackage{hyperref}
\usepackage{breakcites}
\usepackage{amsmath,amssymb,amsthm} 
\usepackage{url} 
\usepackage{float}
\usepackage{graphicx}
\usepackage{subfigure}
\usepackage{color}
\usepackage{lineno}
\usepackage{enumitem}  
\usepackage{chngcntr}
\newcommand{\eqb}{\begin{eqnarray}}
\newcommand{\eqe}{\end{eqnarray}}
\newcommand{\bi}{\begin{itemize}}
\newcommand{\ei}{\end{itemize}}

\newcommand{\txs}{TXS~0506+056}
\newcommand{\icv}{IceCube-200107A}
\newcommand{\hsp}{3HSP~J095507.9+355101}

\newcommand{\swift}{\emph{Swift}}
\newcommand{\fermi}{\emph{Fermi}}

\newcommand{\dop}{\mathcal{D}}

\submitjournal{ApJ}

\shorttitle{Hadronic Blazar Flares}
\shortauthors{Mastichiadis \& Petropoulou}
\graphicspath{{./}{figures/}}

\begin{document}

\title{Hadronic X-ray Flares from Blazars}

\correspondingauthor{Apostolos Mastichiadis}
\email{amastich@phys.uoa.gr}
\correspondingauthor{Maria Petropoulou}
\email{mpetropo@phys.uoa.gr}

\author{Apostolos Mastichiadis}
\affil{Department of Physics, National and Kapodistrian University of Athens, Panepistimiopolis, GR 15783 Zografos, Greece}

\author[0000-0001-6640-0179]{Maria Petropoulou}
\affil{Department of Astrophysical Sciences, Princeton University,
Princeton, NJ 08544, USA}
\affil{Department of Physics, National and Kapodistrian University of Athens, Panepistimiopolis, GR 15783 Zografos, Greece}


\begin{abstract}
The detection of a high-energy neutrino from the flaring blazar TXS 0506+056 and the subsequent discovery of a neutrino excess from the same direction have strengthened the hypothesis that blazars are cosmic neutrino sources. The lack, however, of $\gamma$-ray flaring activity during the latter period challenges the standard scenario of correlated $\gamma$-ray and high-energy neutrino emission in blazars. We propose instead that TeV--PeV neutrinos are produced in coincidence with X-ray flares that are powered by proton synchrotron radiation. In this case, neutrinos are produced by photomeson interactions of protons with their own synchrotron radiation, while MeV to GeV $\gamma$-rays are the result of synchrotron-dominated electromagnetic cascades developed in the source.  Using a time-dependent approach, we find that this ``pure hadronic
flaring'' hypothesis has several interesting consequences. The X-ray flux is a good proxy for the all-flavor neutrino flux, while certain neutrino-rich X-ray flares may be dark in GeV--TeV $\gamma$-rays. Lastly, hadronic X-ray flares are accompanied by an equally bright MeV component that is detectable by proposed missions like \emph{e-ASTROGAM} and \emph{AMEGO}.  We then applied this scenario to the extreme blazar 3HSP~J095507.9+355101 that has been associated
with IceCube-200107A while undergoing an X-ray flare. We showed that the number of muon and antimuon neutrinos above 100~TeV during hadronic flares can be up to $\sim3-10$ times higher than the expected number in standard leptohadronic models. Still, frequent hadronic flaring activity is necessary for explaining the detected neutrino event IceCube-200107A.  
\end{abstract}

\keywords{galaxies: active --- %
  gamma-rays: galaxies --- %
  neutrinos --- %
  radiation mechanisms: non-thermal}

\section{Introduction}\label{sec:intro}
 In 2013 the IceCube neutrino telescope at the South Pole discovered a diffuse and isotropic flux of neutrinos of astrophysical origin \citep{Aartsen:2013a, Aartsen:2013b}. In 2017, the detection of IceCube-170922A in spatial and temporal coincidence with a $\gamma$-ray flare from blazar \txs~provided the first $\sim 3\sigma$ neutrino source association at the time \citep{IceCube:2018dnn}. A follow-up archival search at the position of the blazar revealed an excess of high-energy neutrinos with respect to the atmospheric background in 2014/15. This finding provided a $\sim 3.5\sigma$ evidence for neutrino emission from the direction of \txs~\citep{IceCube:2018cha}. Interestingly, 
during that time the source was at a low state in radio, optical and GeV $\gamma$-rays. More recently the extreme blazar \hsp~was also tentatively associated with IceCube-200107A ~\citep{2020GCN.26655....1I} while it was undergoing an X-ray flare \citep{2020GCN.26669....1G,2020ATel13394....1G,2020Atel13395....1K}. Despite the fact that this was also an observation of low significance, it has further strengthened the possibility that blazar jets are sites of high-energy neutrino emission. 

The tentative association of the neutrino sources with blazars revived the so-called hadronic models of blazar emission. These postulate that the produced $\gamma$-rays arise either directly from synchrotron radiation of relativistic protons accelerated in the jets of these objects \citep{2000NewA....5..377A, 2001APh....15..121M} or indirectly from photomeson collisions with  ambient or co-spatial photon fields \citep[e.g.,][]{1991A&A...251..723M,1993A&A...269...67M, 2003APh....18..593M}. Even before the \txs~results, detailed multi-wavelength modeling, which attempts to describe the observed photon spectrum  and  self-consistently predict the neutrino emission, has revealed that there are sources with different hadronic contributions to their GeV $\gamma$-ray flux \citep{Petropoulou2015}. Contrary to expectations, modeling efforts of \txs~have shown that its broadband photon spectrum is explained by synchrotron and Compton processes of an accelerated electron population (i.e., leptonic model). The hadronic component was found to be radiatively sub-dominant, thus lowering considerably the expected neutrino emission \citep[see e.g.,][]{Keivani2018,2019MNRAS.483L..12C, Gao2019}.

In this paper we examine the expected neutrino emission from blazars by assuming that this is produced not in steady state but during flaring events connected to sporadic proton acceleration episodes in blazar jets. These hadronic flares would commence by photons emitted via the proton synchrotron radiation \citep[see also][for a study of  variability in purely hadronic blazar models]{2013MNRAS.434.2684M}. In cases, however, where these photons are sufficiently energetic and reach high enough densities, they could serve as targets for photomeson interactions on the relativistic protons leading to neutrino production.

The electromagnetic signatures of the hadronic flares under study will depend chiefly on the evolution of proton-initiated electromagnetic cascades that ensue from the secondaries produced in photomeson, photopair, and photon-photon pair production processes. While extensive work on proton-initiated cascades in AGN has been done in the nineties \citep[e.g.,][]{1991PhRvL..66.2697S, 1991A&A...251..723M, 1992A&A...253L..21M, 1993A&A...269...67M,1999MNRAS.302..373B}, it focused on the steady-state cascade spectra. However, the cascades take time to develop because the cooling times of the produced lower energy particles become increasingly longer as the particle energy degrades and, most probably, these cascades will not have reached a steady state, if the source undergoes changes which are faster than the particle cooling times. Therefore a time-dependent approach for the study of hadronic flares is required. This is further justified from the inherent non-linearity of the problem, i.e., from the fact that  protons do not interact only with their own synchrotron radiation, but also with the radiation produced through the cascades.

Here we focus on X-ray flares powered by  proton synchrotron radiation. We have chosen to investigate this regime because it proves to be the most promising for neutrino production: X-rays are energetic targets for photomeson interactions to occur in collisions with relativistic protons and, at the same time, they can be plentiful providing substantial optical thickness for the interactions. Additional motivation is provided by the fact  that X-ray photons are not usually considered as targets for neutrino production in blazars  while proton-induced emission is rarely invoked to explain the low-energy hump of the blazar SED.

The paper is structured as follows. In Section~\ref{sec:analytic} we present  analytical estimates of the necessary conditions required to produce neutrinos in a hadronic flare, and approximate scaling relations between the neutrino and photon luminosities. In Section~\ref{sec:numerical} we describe the numerical approach used to simulate hadronic X-ray flares. In Section~\ref{sec:results} we present our results of saturated hadronic X-ray flares in which the neutrino and photon luminosities are comparable. In Section~\ref{sec:extreme} we apply our results to the case of the extreme blazar \hsp \, that has been tentatively associated with the neutrino \icv~during an X-ray flare. In Section~\ref{sec:future} we discuss the prospects of detecting such hadronic flares in hard X-rays and $\gamma$-rays with future observatories. In Section~\ref{sec:discussion} we discuss our main results, and conclude in Section~\ref{sec:conclusion}.

\section{Analytical considerations}\label{sec:analytic}
In this Section we use qualitative arguments to explain the way the neutrino luminosity relates to the bolometric photon luminosity for various proton cooling regimes and define saturated hadronic flares in the context of the present work. Furthermore,  we derive constraints on the magnetic field strength of the source and the proton energy making use of the appropriate energy thresholds for proton-photon interactions \citep[see e.g.,][]{1990ApJ...362...38B, 1994A&A...286..983M}.

Relativistic protons with Lorentz factor $\gamma'_p$ in the presence of a magnetic field with strength $B'$ radiate synchrotron photons of energy (in $m_e c^2$ units) $x_{p,\rm syn}=b \gamma^{'2}_p (m_e/m_p)$, where $b\equiv B'/B_{\rm cr}$ and $B_{\rm cr}=4.4\times10^{13}$~G. We can  express $x_{p,\rm syn}$  in terms of the photon energy in the observer's frame, $\varepsilon_{ph, \rm keV}\equiv \varepsilon_{ph}/1~{\rm keV}$, and the Doppler factor of the emission region, $\dop$, as
\eqb 
B' = 16 \,  \varepsilon_{ph, \rm keV} \dop_1^{-1} \gamma^{'-2}_{p,6}~{\rm G},
\label{eq:xpsyn}
\eqe 
where we introduced the notation $\mathcal{D}_1\equiv \mathcal{D}/10$ and $\gamma'_{p,6}\equiv \gamma'_p/10^6$. For simplicity, we did not include the $1+z$ term in the equation above. Primed quantities are measured in the rest frame of the emission region, while unprimed quantities are measured in the observer's frame. 

Protons can also interact via photopair (Bethe-Heitler) and photomeson production processes on their own synchrotron radiation \citep[for illustrative examples, see][]{2012A&A...546A.120D}. The energy threshold condition for photopair production on proton synchrotron photons reads $x_{p, \rm syn} \gamma'_p \gtrsim 2$ or
\eqb 
B'\gtrsim 0.16\, \gamma^{'-3}_{p,6}~\rm G,
\label{eq:BH-thr-psyn}
\eqe 
and the energy threshold condition for photomeson production on proton synchrotron photons reads $x_{p, \rm syn} \gamma'_p \gtrsim (m_\pi/m_e)(1+m_\pi/2m_p)$ or 
\eqb 
B' \gtrsim  24 \, \gamma^{'-3}_{p,6}~\rm G.
\label{eq:pg-thr-psyn}
\eqe 
Even if the latter condition is not satisfied simultaneously with equation~(\ref{eq:xpsyn}), protons can still pion-produce on higher energy photons that are available in the source. These can be synchrotron photons from Bethe-Heitler pairs with typical energy  $x_{e, \rm syn}=b \gamma^{'2}_e \approx b \gamma^{'2}_p$. The energy threshold condition for photomeson production is relaxed by a factor of $m_e/m_p$ with respect to that given by equation~(\ref{eq:pg-thr-psyn}), namely
\eqb 
 B' \gtrsim 1.2\times10^{-2}\, \gamma^{'-3}_{p,6}~\rm G.
\label{eq:pg-thr-BHsyn}
\eqe 

\begin{figure}
    \centering
    \includegraphics[width=0.49\textwidth]{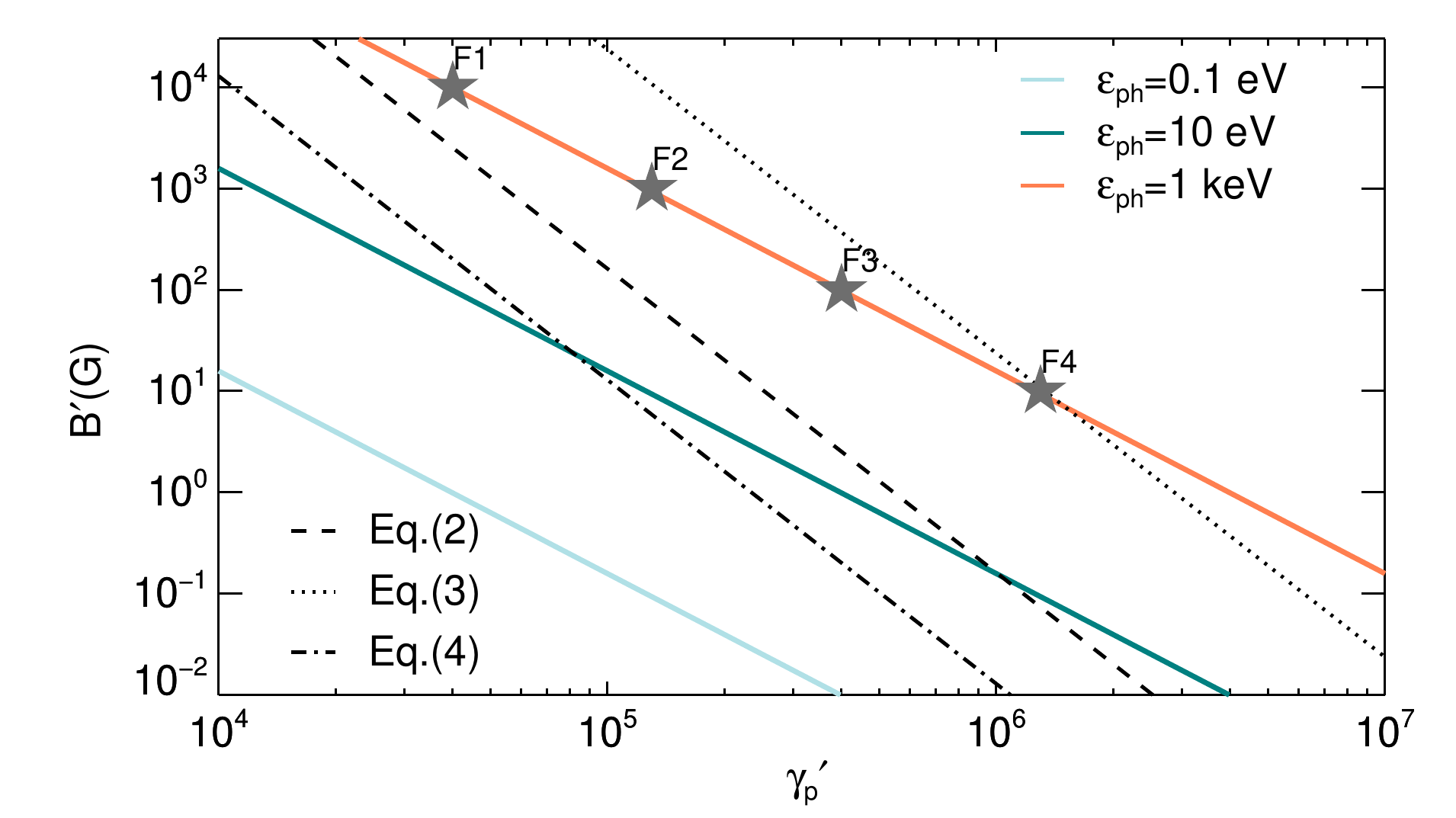}
    \caption{Threshold conditions for photopair and photomeson interactions in a hadronic flare plotted as a function of the proton Lorentz factor $\gamma'_p$ and the magnetic field of the source $B'$. 
    The threshold for photopair and photomeson production on proton synchrotron photons is satisfied above the dashed and dotted lines, respectively. The threshold for photomeson production on synchrotron photons from Bethe-Heitler pairs is satisfied above the dot-dashed line.
    Solid colored lines indicate combinations of $B', \gamma'_p$ that yield proton synchrotron photons of energy  0.1 eV (light blue line), 10 eV (teal line) and 1~keV (orange line). The parameters for the numerical models discussed in Section~\ref{sec:results} are indicated with stars.}
    \label{fig1}
\end{figure}

The conditions given by equations (\ref{eq:xpsyn})-(\ref{eq:pg-thr-BHsyn}) are displayed in Figure~\ref{fig1} for $\dop=10$. If
$\varepsilon_{p, syn} \equiv \mathcal{D} x_{p,syn} m_e c^2=0.1$~eV 
(solid light blue line), then neither photopair nor photomeson production on the synchrotron photons can occur, and no high-energy neutrino and photon emission is expected. If $\varepsilon_{p, syn}=10$~eV proton synchrotron photons are more energetic 
(solid teal line), then there is a combination of $B', \gamma'_p$ values that satisfies the threshold for photopair production (dashed line). Even though photomeson production on proton synchrotron photons is not possible (solid teal line lies below the dotted line), protons can still pion produce on the Bethe-Heitler synchrotron photons (solid teal line lies above the dot-dashed line). In this case, $L'_{\nu} \propto L^{'3}_{p,syn}$, where $L'_{\nu}$ is the bolometric neutrino and antineutrino luminosity, and $L'_{p,syn}$ is proton-synchrotron luminosity. If $\varepsilon_{p, syn}=1$~keV
(solid orange line), it is possible that protons can interact with their own synchrotron photons to produce pions (i.e., solid orange line lies above the dotted line). Although synchrotron photons from Bethe-Heitler pairs are an additional target for photomeson interactions,  proton synchrotron photons dominate the pion production due to their higher number density. In this case, $L'_{\nu} \propto L^{'2}_{p,syn}$.

For the purposes of this qualitative discussion, let us consider the case of mono-energetic protons. The bolometric photon luminosity produced by relativistic protons that are injected with luminosity $L'_p$ in the source can be written as
\eqb
L'_{ph} = \frac
{t^{'-1}_{\rm syn} + t^{'-1}_{\rm BH} + \alpha t^{'-1}_{\rm mes}} {t^{'-1}_{\rm syn} + t^{'-1}_{\rm BH} + t^{'-1}_{\rm mes} + t^{'-1}_{\rm esc}}L'_{p},
\label{eq:Lg}
\eqe 
where $t^{'-1}_{\rm syn}$, $t^{'-1}_{\rm BH}$ and $t^{'-1}_{\rm mes}$ are the proton synchrotron, Bethe-Heitler and photomeson energy loss rates, respectively,
while $t^{'-1}_{\rm esc}$ is the proton physical escape rate from the source. The above relation is valid when secondary electron-positron pairs radiate very fast their energy. Here, $\alpha\simeq 5/8$ is the fraction of proton energy that goes to photons and electron-positron pairs in a photomeson collision, while the rest is assumed to go to neutrinos and antineutrinos. 

The all-flavor neutrino luminosity is then written as   
\eqb
L'_\nu=\frac{(1-\alpha)t^{'-1}_{\rm mes} } {t^{'-1}_{\rm syn} + t^{'-1}_{\rm BH} + t^{'-1}_{\rm mes} + t^{'-1}_{\rm esc}}L'_{p}.
\label{eq:Lv}
\eqe 

The above equations provide a rough guide for the relation between the photon and neutrino luminosity, which can be applied also to the case of extended proton energy distributions. Assuming that $t^{'-1}_{\rm BH} \ll t^{'-1}_{\rm mes}$  \citep[but see also][]{2015MNRAS.447...36P}, we can ignore Bethe-Heitler pair production and distinguish the following cases according to the efficiency of each process
\begin{enumerate}[label=(\roman*)]
\item If  $t^{'-1}_{\rm mes} \ll  t^{'-1}_{\rm syn} \ll  t^{'-1}_{\rm esc}$, then $L_\nu \ll L_{ph} \ll L_{p}$.
\item If $t^{'-1}_{\rm mes} \ll t^{'-1}_{\rm esc} \ll t^{'-1}_{\rm syn}$, then $L_\nu \ll L_{ph} \lesssim L_{p}$.
\item If $t^{'-1}_{\rm esc} \ll t^{'-1}_{\rm mes} \simeq t^{'-1}_{\rm syn}$, then $L_\nu \simeq L_{ph} \lesssim L_{p}$.
\end{enumerate}

Case (i) corresponds to a slow cooling regime which is characterized by low efficiency in both photon and neutrino production. Case (ii) indicates efficient photon production through synchrotron, but not through photomeson processes, thus leading to much lower neutrino output compared to the bolometric electromagnetic output of the source. Only case (iii) can lead to efficient neutrino production, thus raising the question of the required conditions under which the relation $t^{'-1}_{\rm mes} \simeq t^{'-1}_{\rm syn}$ holds. We do not consider cases with $t^{'-1}_{\rm mes} \gg t^{'-1}_{\rm syn}$, as these usually lead to proton-photon runaways with heavily modified photon spectra 
\citep[e.g.,][]{1992Natur.360..135K, 2014MNRAS.444.2186P, 2020MNRAS.495.2458M}, whose applicability to blazar emission requires a dedicated study.

Let us take a closer look at the ``neutrino-rich'' case. Dividing by parts equations (\ref{eq:Lg}) and (\ref{eq:Lv}), we derive the useful relation
\eqb
\frac{L_\nu}{L_{ph}}=\frac{(1-\alpha)t^{'-1}_{\rm mes}}{t^{'-1}_{\rm syn}+\alpha t^{'-1}_{\rm mes}}
\simeq\frac{3}{8\xi+5}
\eqe 
where $\xi=t^{'-1}_{\rm syn}/t^{'-1}_{\rm mes}$.
The neutrino luminosity starts increasing for decreasing $\xi$. Note, however, that even when $\xi\rightarrow 0$ (that applies to proton-photon runaway systems) the ratio of the luminosities approaches the constant value of 0.6, which is dictated by the branching ratio in photomeson interactions. Therefore, the neutrino luminosity will be, at best, of the same order as the photon luminosity. This means that the photon and neutrino components will be approximately at the same flux level in a $\varepsilon F_\varepsilon$ plot.

In the absence of external photon fields, as assumed here, all target photons for photomeson production are internally produced. These consist mainly of proton-synchrotron photons and synchrotron photons from their secondaries\footnote{Inverse Compton scattering is largely suppressed due to Klein-Nishina effects.}.  On the one hand, $t^{'-1}_{\rm syn}\propto U'_{\rm B}$, where $U'_{\rm B}=B^{'2}/8\pi$. On the other hand,
$t^{'-1}_{\rm mes}\propto U'_{ph}$ with $U'_{ph}$ being the  co-moving photon energy density above the relevant energy threshold that depends on both $B'$  and the number density of relativistic protons. 
The two rates become comparable if the source is compact, i.e., has a large ratio $L'_{ph}/R'$. This implies that there is always a value of $L'_p$ that can lead to $\xi \approx 1$ in a source of given $R'$ and $B'$, thus maximizing the expected neutrino emission of a hadronic X-ray flare. Henceforth, we refer to the latter as \emph{saturated} flares. 

Although the considerations presented so far
are useful for a qualitative understanding of the multi-messenger behavior of a hadronic flare, they cannot provide information about the shape or time-dependence of the photon and neutrino fluxes. This is because the broadband photon spectrum is shaped by non-linear processes that cannot be tracked analytically, such as electromagnetic cascades initiated by intra-source photon-photon absorption. In addition, the ratio $\xi$ is expected to vary during a flare, and might also drive the system into a non-linear cooling regime, where protons start losing energy on the radiation of their own secondaries. For these reasons, we proceed in the next section with a numerical study of hadronic flares.

\section{Numerical approach}\label{sec:numerical}
We describe our approach for the numerical study of hadronic flares that are neutrino-rich and have a proton synchrotron peak in X-rays. More specifically, we assume that the peak energy of the flare occurs at $\varepsilon_{ph}=1$~keV and the flare reaches a peak luminosity $L_{ph}\sim10^{45}$~erg s$^{-1}$.  

We use the numerical code {\sc athe$\nu$a} \citep{1995A&A...295..613M, 2012A&A...546A.120D} that computes the multiwavelength photon and all-flavor neutrino spectra as a function of time by solving the kinetic equations for relativistic protons, secondary electrons and positrons, photons, neutrons, and neutrinos. 

We assume that prior to the onset of a hadronic flare, the emitting region contains a very low number of relativistic particles whose emission can be neglected. We then consider that relativistic protons are injected into the emission region, which is modeled as a spherical blob of radius $R'$ with magnetic field $B'$ and Doppler factor $\dop=10$. The proton distribution at injection is taken to be a power law with slope $p$ extending from $\gamma'_{p,\min}$ and $\gamma'_{p,\max}$.
Here we fix $\gamma'_{p,\min}=1$ assuming that acceleration starts from particles essentially at rest. Under this assumption the flux of secondaries produced in the photomeson production process generally increases with decreasing $p$, since more power is carried by protons with higher energies relevant for neutrino production \citep[see e.g., Figure 12 in][]{2012A&A...546A.120D}. To minimize the energetic requirements of the hadronic flare, we therefore consider hard power laws ($p<2$). In this case, the proton maximum energy is the main free parameter of the model. Without loss of generality we set $p=1.5$.

To simulate a hadronic flare, we vary the total injection proton luminosity, $L'_p$, according to a Lorentzian temporal profile
\eqb 
L'_p(t')=L'_{p,\rm pk} \frac{w_p^{'2}}{w_p^{'2}+4(t'-t'_{\rm pk})^2} \cdot
\label{eq:Lp}
\eqe 
Here, $L'_{p, \rm pk}$ is the value of the proton luminosity which occurs at time $t'_{\rm pk}$, and $w'_p$ is the time interval where $L'_p> 0.5 \, L'_{p,\rm pk}$. The pulse profile of the induced photon flares can also be described by equation (\ref{eq:Lp}). However, the exact values of the peak luminosity and full width at half maximum (FWHM) will depend on the selected energy band, and on the proton radiative efficiency.
 
The requirement that proton synchrotron radiation produces a flare in X-rays constrains $B' \gamma_{p,\max}^{'2}$ (see equation \ref{eq:xpsyn}). For the numerical study of hadronic X-ray flares, we select four pairs of $B', \gamma'_{p,\max}$ values marked with symbols in Figure~\ref{fig1}. We assume that $B' \propto R^{'-1}$, thus minimizing the number of free parameters. It turns out that the choice of this scaling relation is not crucial for the derivation of our conclusions. By fixing the peak luminosity of the X-ray flare, we can 
also constrain $L'_{p,\rm pk}$. The FWHM of the proton injection pulse profile, $w'_p$, is chosen so that FWHM of the X-ray flare in the observer's frame is around one day. However, our main conclusions are not sensitive on the value of $w'_p$.  Finally, we present an additional case (Flare 5) that will be discussed  in application to the X-ray flare of the extreme blazar \hsp \, see Section~\ref{sec:extreme}). Table~\ref{tab:param} summarizes the parameter values used for the simulation of the hadronic flares.

\begin{deluxetable}{ccc ccc}
\centering
\tablecaption{Parameter values used for the numerical study of hadronic X-ray flares.\label{tab:param}}
\tablewidth{0pt}
 \tablehead{
 \colhead{Flare} & \colhead{$R'$ (cm)}  & \colhead{$B'$ (G)}  & \colhead{$\gamma'_{p,\max}$}  & 
 \colhead{$w'_p$ (d)}  & \colhead{$L'_{p, \rm pk}$ (erg s$^{-1}$)}  
 }
 \startdata
 1 & $10^{13}$ & $10^4$ & $10^{4.7}$ & 19   & $2.2\times10^{45}$ \\
 2 &$10^{14}$ & $10^3$ & $10^{5.2}$ &10    & $8.6\times10^{45}$  \\
 3 &$10^{15}$ & $10^2$ & $10^{5.7}$ & 10   & $3.8\times10^{46}$\\
 4 &$10^{16}$ & $10$ & $10^{6.2}$ & 10   & $2.2\times10^{47}$ \\
 5 & $4\times10^{14}$ & $2.5\times10^2$ & $10^{5.5}$ & 60 & $2.0\times10^{46}$  \\
\enddata
\tablecomments{Flare 5 is introduced in Section~\ref{sec:extreme} as application to the X-ray flare of the extreme blazar \hsp.}
\end{deluxetable} 

\section{Results}\label{sec:results}
\begin{figure*}
    \centering
    \includegraphics[width=0.48\textwidth]{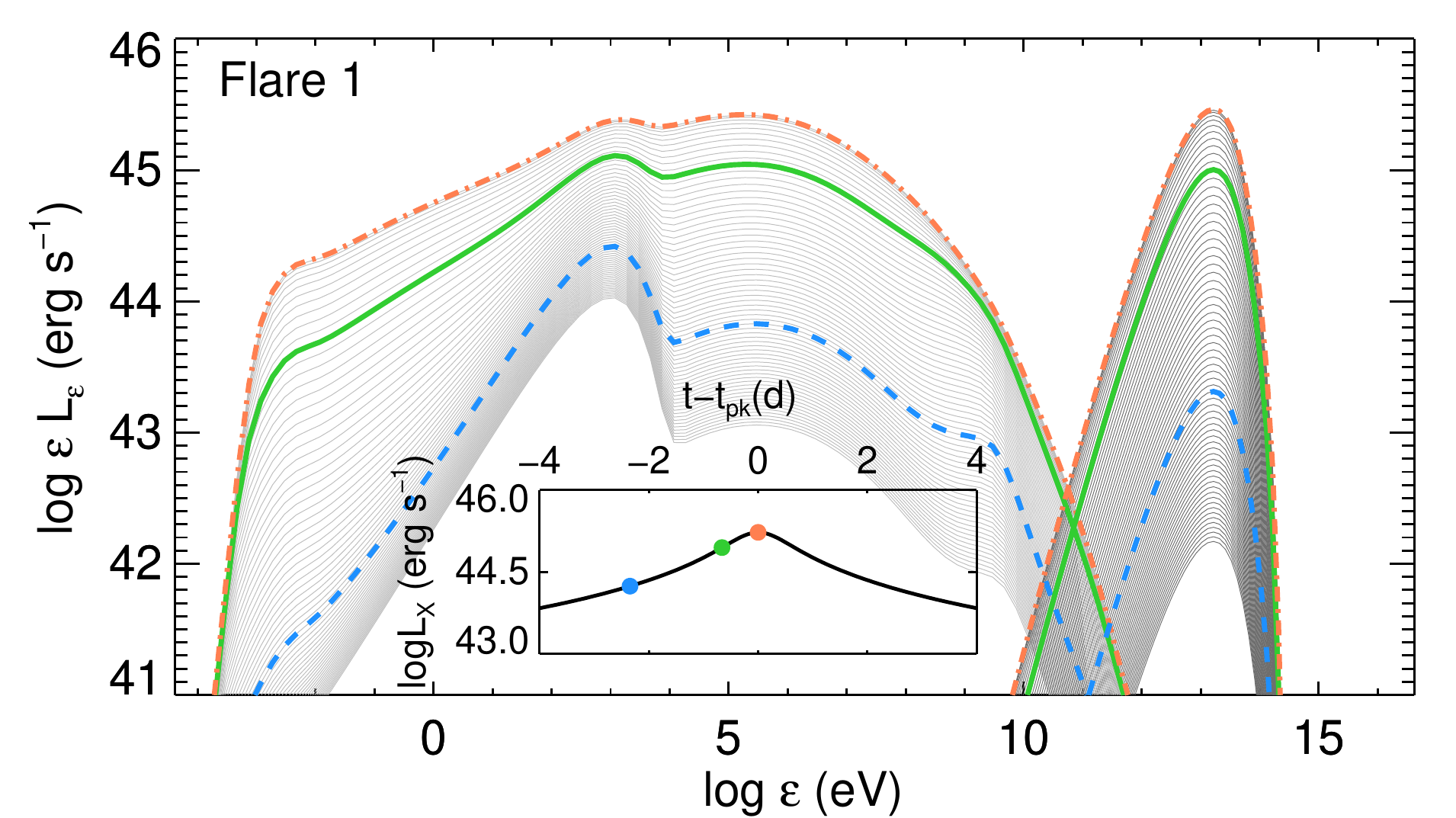}
    \includegraphics[width=0.48\textwidth]{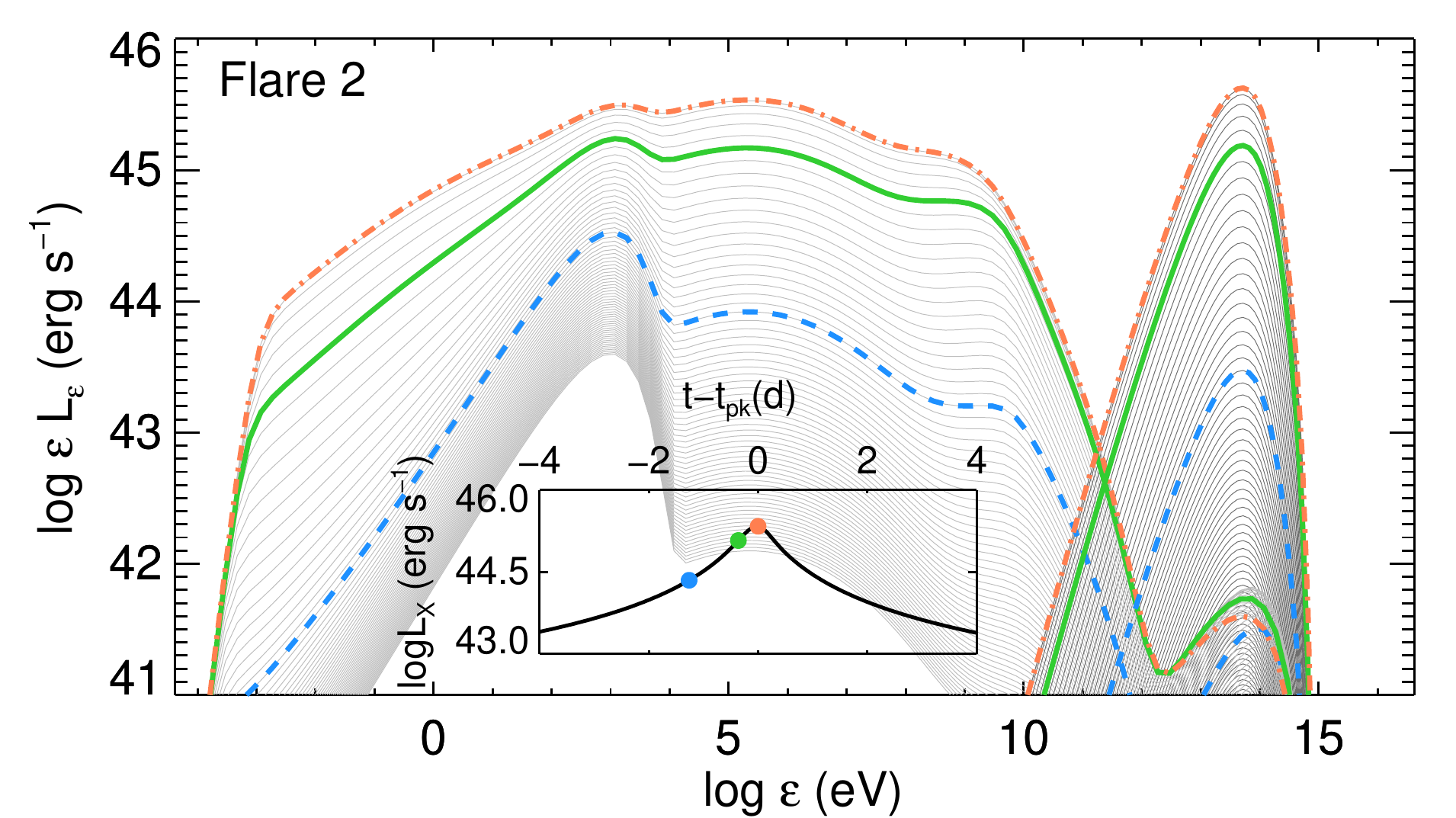}
    \includegraphics[width=0.48\textwidth]{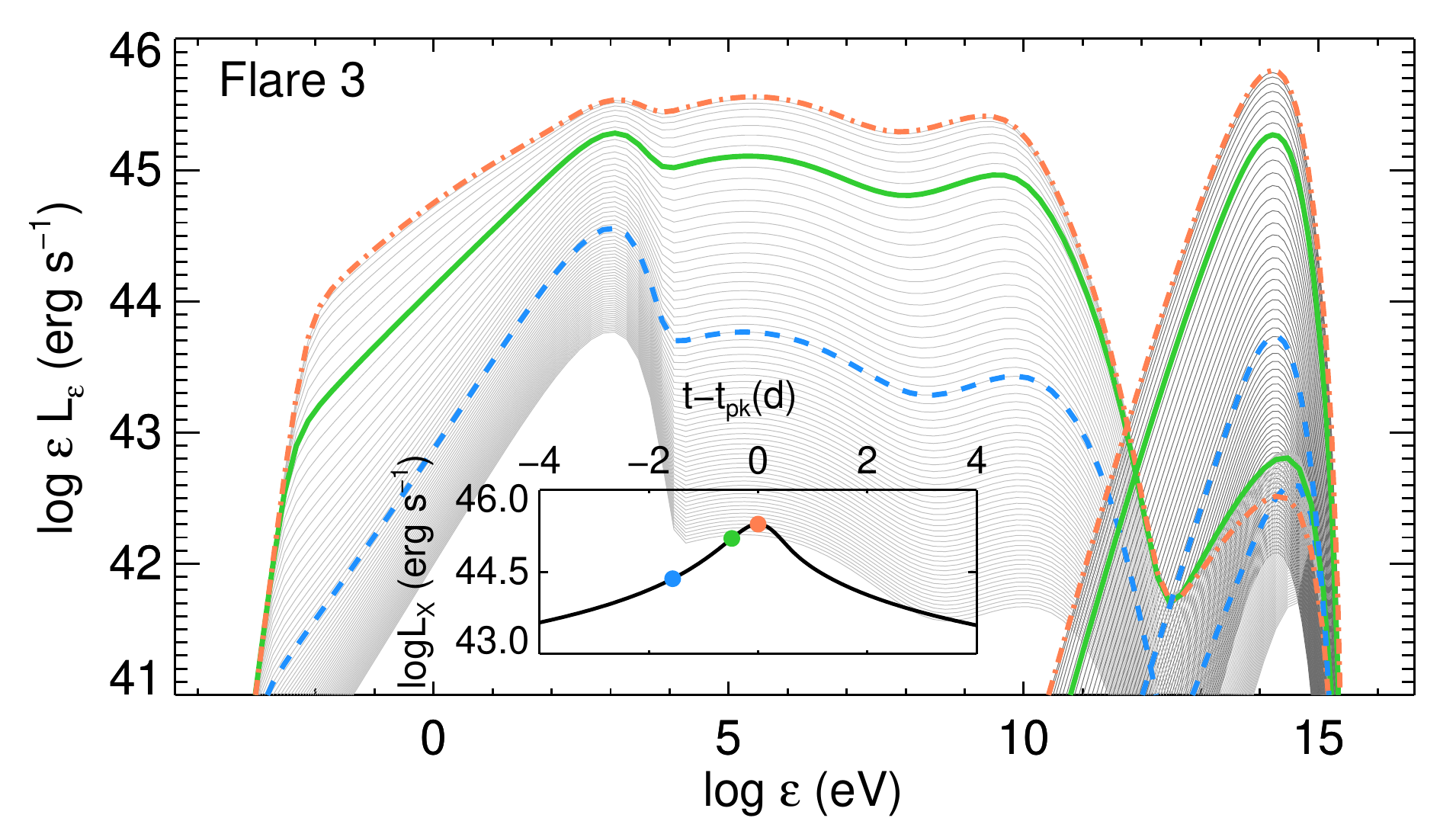} 
    \includegraphics[width=0.48\textwidth]{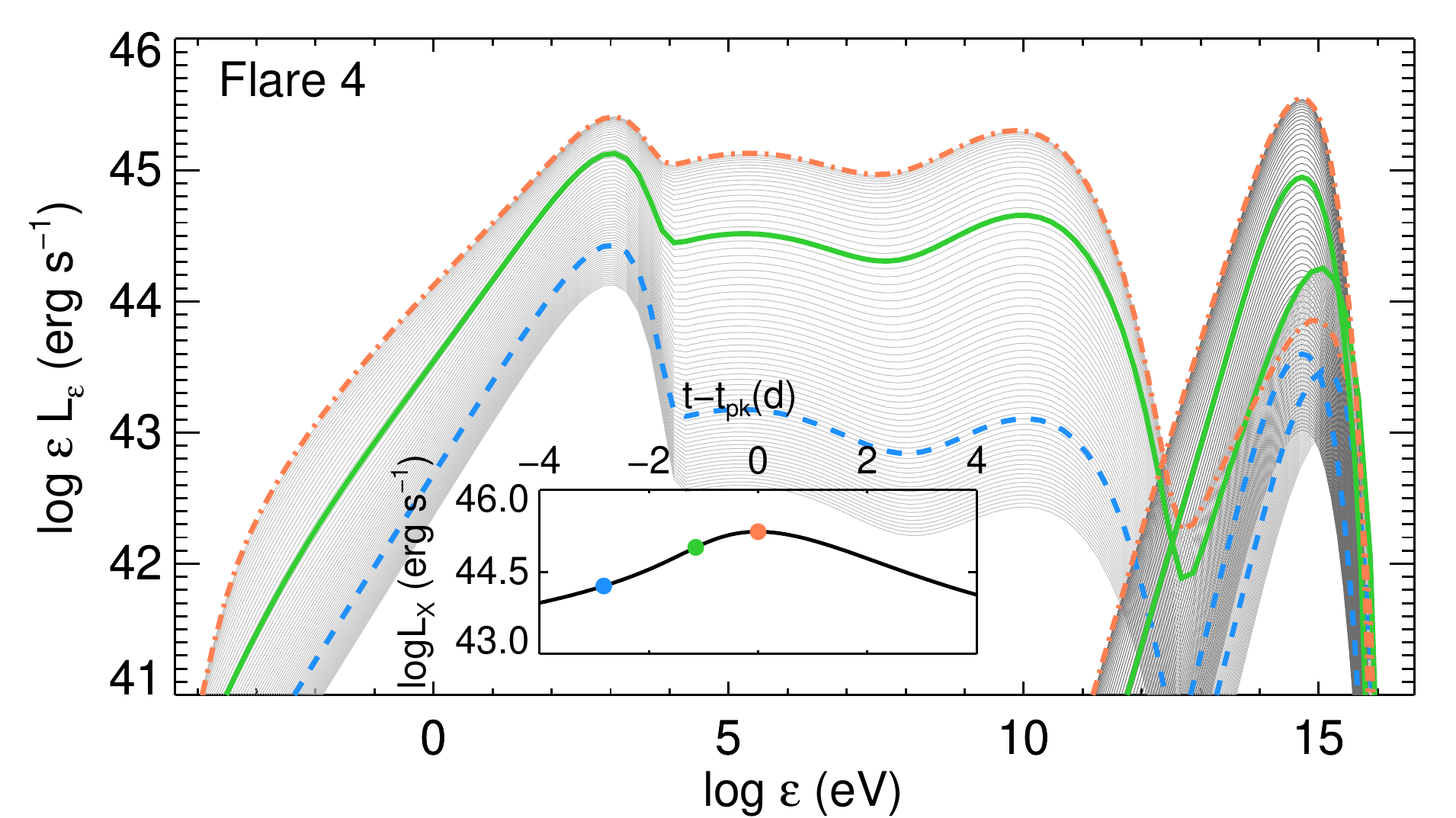}  
    \caption{Temporal evolution of photon (light grey lines) and all-flavor neutrino (dark grey lines) energy spectra for four indicative hadronic X-ray flares (for their parameters, see Table~\ref{tab:param} under Flares 1 to 4).  Solid grey lines show snapshots every $\sim1.5$~hr in the observer's frame for a 4-day time window prior to the  peak of the X-ray flare. Colored lines show snapshots corresponding to 10\% (dashed line), 50\% (solid line), and 100\% (dash-dotted line) of the 1-2~keV flare's peak luminosity. Inset plots in all panels show a zoom in the 1-2~keV light curve around the peak time of the flare.}
    \label{fig2}
\end{figure*}
Figure~\ref{fig2} shows our results for four exemplifying hadronic X-ray flares, and their accompanying neutrino emission, computed using the methods described in the previous section; the characteristic sets of parameters used are listed in Table~\ref{tab:param}. Despite the widely different parameter values, all ensued flares have some features in common.

At early times (see e.g., blue dashed lines) the bolometric luminosity is still low, the ratio $\xi >1$, and the synchrotron component, which peaks at 1~keV, dominates the total emission. The electromagnetic emission from photohadronic interactions, which consists mostly of synchrotron radiation of secondaries, is $\sim10$ times less luminous than that of the proton synchrotron component. The same can be said about the neutrino emission. As the luminosity increases, $\xi$ is decreasing, and the photohadronic emission accounts for a significant fraction of the bolometric photon luminosity  (see grey lines between dashed blue and solid green lines). Close to the peak (solid green lines) and at the peak (dot-dashed orange lines) of the flare, the multiwavelength photon spectrum is characterized by strong photohadronic emission. The all-flavor neutrino emission is at the same flux level (in $\varepsilon F_{\varepsilon}$ units) as the photon emission. This is the highest possible expected neutrino flux for the selected parameters (i.e., $B'$, $\gamma'_{p,\max}$, and $R'$), as explained in  Section~\ref{sec:analytic}. \emph{Therefore, the energy flux of a saturated hadronic X-ray flare is a good proxy for the all-flavor neutrino energy flux.} 

Another similarity is that all photon spectra (close to the peak time of the flare) have a broad spectral component in soft gamma-rays ($\sim 0.1-100$~MeV), whose flux is comparable to the X-ray flux and the all-flavor neutrino flux. This spectral feature arises due to the synchrotron emission of secondary pairs produced directly via photohadronic interactions and indirectly via photon-photon pair production. \emph{The (sub-)MeV emission of a saturated hadronic flare is therefore a second proxy for high-energy neutrino emission.}

The four cases also exhibit some apparent differences. The source size differs by a factor of $10^4$ from Flare 1 to Flare 4 (see Table~\ref{tab:param}), while the radiated luminosity at X-rays and the Doppler factor are kept constant. As a result, the source photon compactness, $L_{ph}/\dop^4 R'$, also varies by four orders of magnitude, with the source of Flare 1 being the most compact. Because the photon compactness is a good measure of the optical depth to photon-photon absorption, the attenuation of $\gamma$-ray photons is strongest in Flare 1, and gradually decreases as we move to Flare 4. This can be also verified visually, when comparing the photon spectra at the peak of each flare. The $\gamma$-ray spectrum of Flare 1 is absorbed above $\sim10$~MeV, while the spectrum of Flare 4,  on the other end, extends unaffected by intra-source attenuation up to $\sim10$ GeV. {\sl Thus, not all neutrino-rich hadronic X-ray flares are strong $\gamma$-ray emitters.}

The contribution of secondaries to the photon emission is not always limited in the soft $\gamma$-ray band and beyond. For flares produced in sufficiently compact sources (e.g., Flares 1-3) the optically thin synchrotron emission of fast-cooling secondary electrons and positrons can dominate the flux below X-rays. At early times, the emission at energies below 1~keV is  solely due to proton synchrotron radiation in all flares. The optically thin spectrum can be described by a power law,  $L(\varepsilon)\propto \varepsilon^{-\beta}$, with index $\beta=(p-1)/2 =0.25$.  The change of the spectral index from its early-time value to $0.5$, which is observed in Flares 1 to 3, indicates a change in the cooling regime of the radiating particles. This in turn implies that the emission is dominated by fast-cooling pairs, which are produced in photohadronic interactions and through photon-photon absorption. If the source is not compact enough, like in Flare 4, the secondary production is suppressed. Together with the lower cooling rate due to the weaker magnetic field and larger source size, the contribution of secondaries to the low-energy flare emission is less important. \emph{Thus, saturated hadronic X-ray flares are always accompanied by optical flares, but the exact correlation of the optical and X-ray fluxes close to the peak time of the flare depends on the source parameters.}

The peak neutrino flux of saturated hadronic X-ray flares is similar to the peak flux reached in X-rays. However, the peak neutrino energy $\varepsilon_{\nu, \rm pk}$ (i.e., the energy where the neutrino spectrum peaks in $\varepsilon F_{\varepsilon}$ units) depends on the source parameters. More specifically, neutrino flares produced by more energetic protons (and thereby weaker magnetic fields, see also Figure~\ref{fig1}) will have higher peak energies. Flare 4, for example, which has a combination of the highest $\gamma'_{p,\max}$ and lowest $B'$ will produce the highest peak neutrino energy with a peak roughly at $\varepsilon_{\nu, \rm pk}\simeq 2~\dop_1 \gamma'_{p,\max}$~GeV \citep{2012A&A...546A.120D} or  $\varepsilon_{\nu, \rm pk}\simeq 2~\varepsilon_{ph, \rm keV}  B^{'-1}_{1.2}  \gamma^{'-1}_{p,6}$~PeV, where we used equation (\ref{eq:xpsyn}).

\begin{figure*}
    \centering
    \includegraphics[width=0.9\textwidth, trim = 20 0 0 0 ]{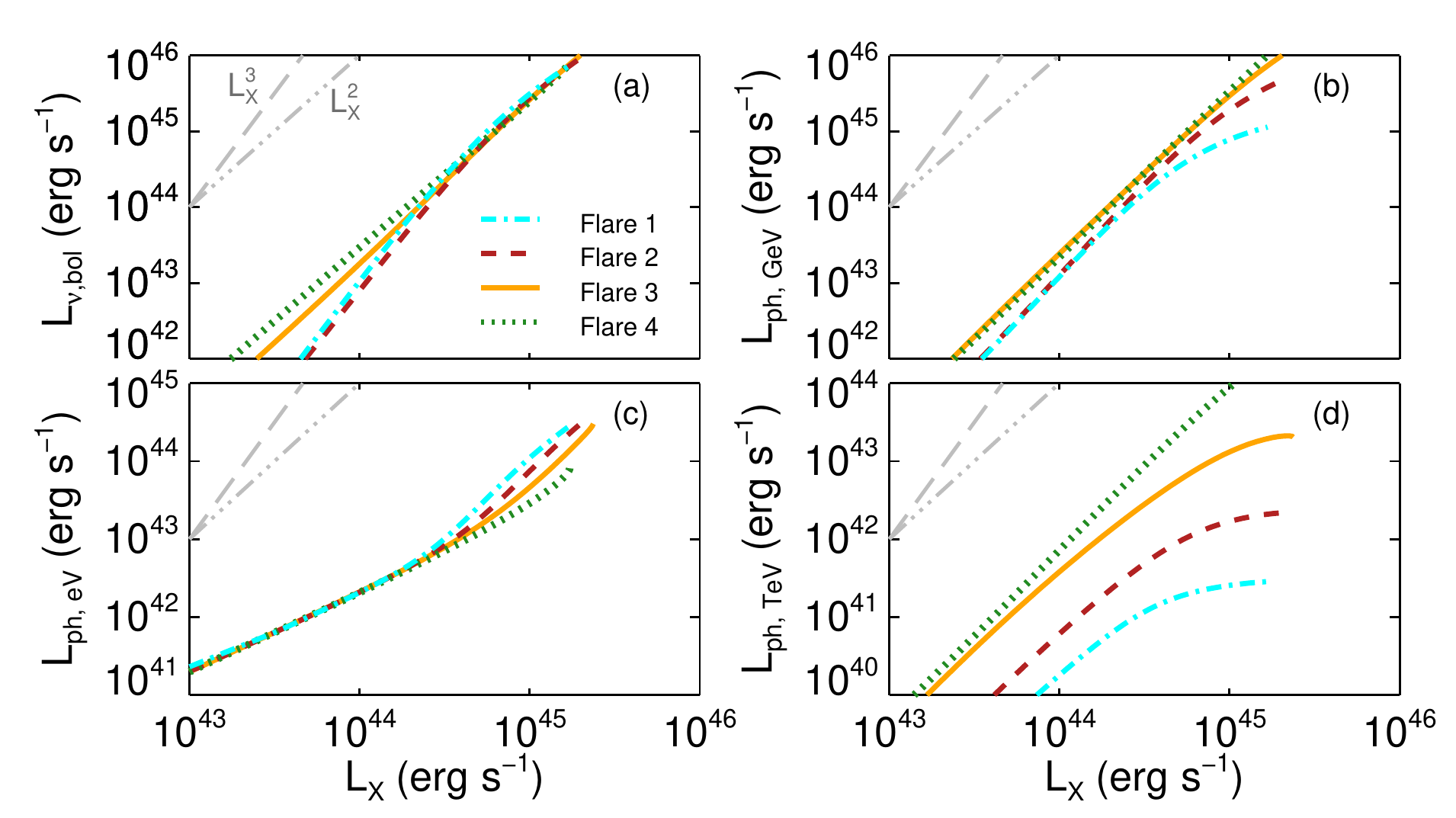}
    \caption{All-flavor bolometric neutrino luminosity ($L_{\nu, bol}$) and photon luminosities in different energy bands ($L_{ph, \rm TeV}$ in 0.3-10 TeV; $L_{ph, \rm GeV}$ in 0.1-300 GeV, and $L_{ph,\rm eV}$ in 1-2 eV) plotted against the X-ray luminosity for the hadronic flares shown in Figure~\ref{fig2} (for a description of the lines, see inset legend). Grey lines with slopes of 2 and 3 (triple-dot-dashed and long-dashed lines, respectively) are plotted to guide the eye.}
    \label{fig3}
\end{figure*}

A more compact view of the various multiwavelength dependencies expected during hadronic X-ray flares is provided in Figure~\ref{fig3}. We show plots of the all-flavor bolometric neutrino luminosity and the photon luminosity
in three characteristic energy bands (see figure caption) versus the X-ray luminosity of Flares 1 to 4.
Flares that are produced in more compact sources (e.g., Flares 1 and 2) tend to have suppressed TeV emission after a certain X-ray luminosity is reached, because of strong  internal photon-photon absorption (panel d). This leads to a saturation of the TeV flux with increasing $L_X$, as indicated by the sub-quadratic scaling of $L_{ph, \rm TeV}$ with $L_X$. In such cases, only before the attenuation becomes important (i.e., at low luminosities), the TeV emission is strongly correlated with the X-ray emission. The same can be said about the GeV emission of the flares, but here the effect of attenuation is less pronounced (panel b). 
Contrary to the $\gamma$-rays, the optical emission is amplified close to the X-ray peak of the flare (panel c). At sufficiently low X-ray luminosities, when the optical flux is dominated by proton synchrotron radiation, we find $L_{ph, \rm eV}\propto L_X$ (panel c). Closer to the X-ray peak of the flare, the optical emission is amplified due to the copious production of secondary pairs produced in photohadronic interactions that are also fast cooling via synchrotron (see also Figure~\ref{fig2}). In this regime, we find $L_{ph, \rm eV}\propto L_X^2$. Thus, the change from a proton-synchrotron to a pair-synchrotron dominated optical emission is indicated by a very specific change in the slope of the curves. Finally, the neutrino luminosity correlates quadratically with the X-ray luminosity close to the peak of the flare (panel a). A cubic relation is found for lower luminosity levels for Flares 1 and 2 that arise from the more compact sources. At the flare maximum, the neutrino luminosity can even exceed the X-ray one by a factor of $\sim3$. This is not in contradiction with our analysis in Section~\ref{sec:analytic}, according to which $L_{\nu} \rightarrow 0.6 L_{ph}$. The bolometric photon  luminosity $L_{ph}$ is spread over many orders of magnitude in frequency due to the intra-source electromagnetic cascades,  thus leading to $L_X < L_{ph}$.

\section{Application to the extreme blazar \hsp}\label{sec:extreme}
In January 2020 IceCube reported the observation of the high-energy neutrino, IceCube-200107A~\citep{2020GCN.26655....1I}. Electromagnetic follow-up of sources within the uncertainty region of the neutrino arrival direction led to the detection of an X-ray flare from the high-synchrotron peaked (HSP) blazar \hsp~\citep{2020GCN.26669....1G,2020ATel13394....1G,2020Atel13395....1K}, which
is part of the 3HSP catalog~\citep{2019A&A...632A..77C}. In fact, with a peak synchrotron frequency of $\varepsilon_s \sim 2$~keV, the source belongs to the rare class of extreme blazars~\citep[for a recent review, see][]{2020NatAs...4..124B}. It has also been detected by the {\it Fermi}-LAT as a $\gamma$-ray emitting source and is thus also included in the 4FGL catalog~\citep{2020ApJS..247...33A}. Subsequent to the detection of the X-ray flare, the redshift of the source was determined to be $z = 0.557$~\citep{2020MNRAS.495L.108P}. 

Here, we make a tentative comparison of a hadronic X-ray flare  to the multi-wavelength data of \hsp~that were obtained within 3 days after the IceCube-200107A detection. For this purpose, we choose Flare 5 (see Table~\ref{tab:param}) and we adopt the multi-wavelength data set presented in Figure 2 of \citet{2020A&A...640L...4G}. 

\begin{figure}
    \centering
    \includegraphics[width=0.47\textwidth, trim = 10 0 0 0]{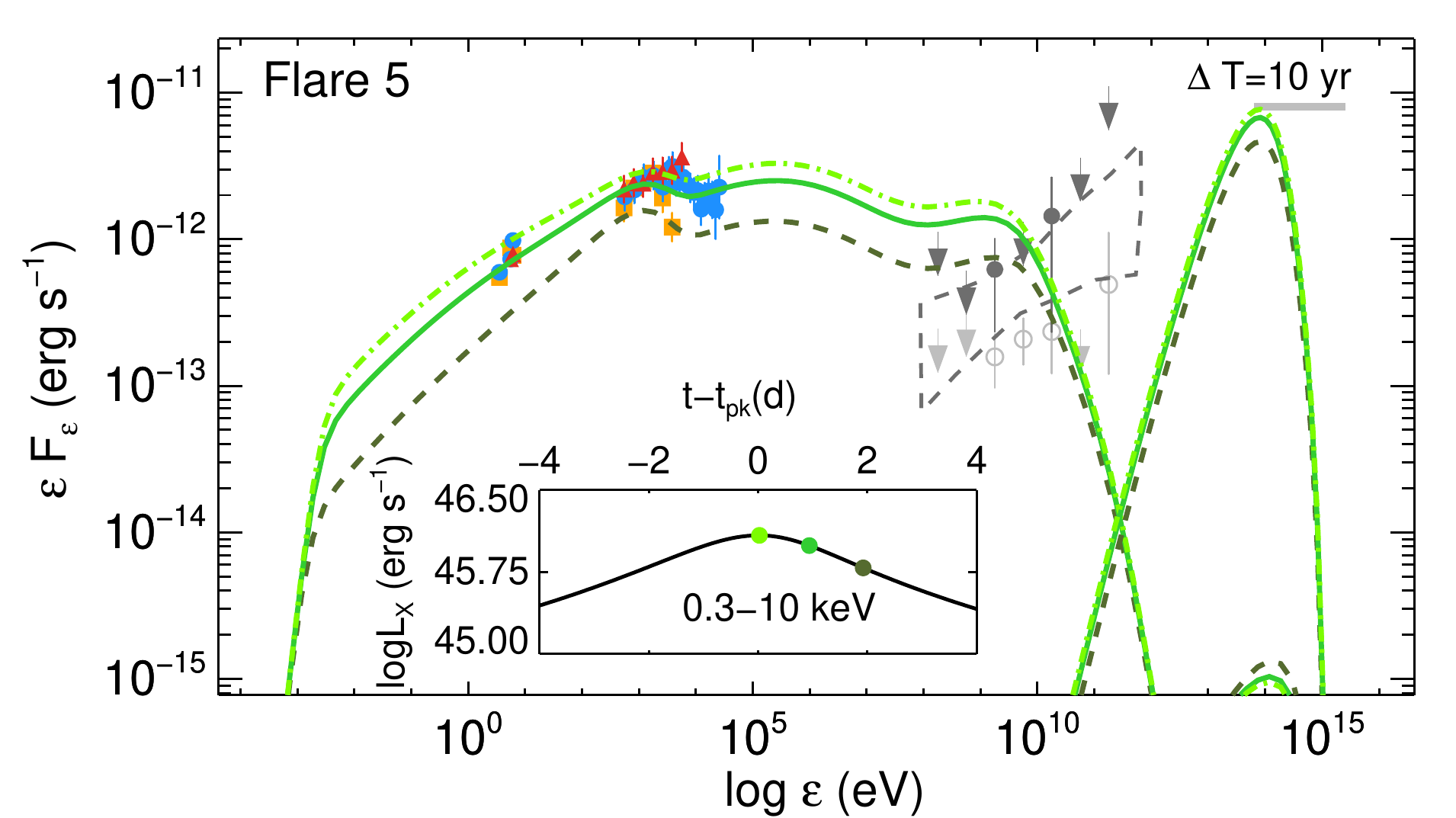} 
    \caption{SEDs of \hsp \ built with data from \citet{2020A&A...640L...4G}. Colored filled symbols indicate observations taken soon after the arrival of the neutrino alert (see inset legend). The inferred all-flavor neutrino flux (assuming an $\varepsilon_\nu^{-2}$ spectrum) is also marked on the plot (horizontal grey lines) for an assumed duration $\Delta T=10$~yr of neutrino emission. The grey bowtie and filled symbols show the time-integrated \fermi-LAT data over a period of 250 days prior to the neutrino alert. Archival data are overplotted with grey open symbols. Colored lines show snapshots of the photon and all-flavor neutrino spectra corresponding to the peak time of the X-ray flare and to two consecutive days post peak. The inset plot shows a zoom in the 0.3--10~keV light curve around the peak time of the flare.}
    \label{fig4}
\end{figure}

Our results are shown in Figure~\ref{fig4}, where the colored lines show snapshots of the photon and all-flavor neutrino spectra corresponding to the peak time of the X-ray flare and to two consecutive days post peak. Given that we did not perform a fit to the data, a disagreement between the model and the data is expected at some level. Nevertheless, for the parameters of Flare 5, we are able to reproduce the optical-to-X-ray spectral index and the synchrotron peak energy flux. A different choice in the temporal profile of the proton injection rate and a time-varying maximum proton energy would alleviate the differences between the model and data in regard to the X-ray spectral cutoff and the optical/X-ray spectrum of the second day post peak (dashed line).

Although the broadband photon spectrum of the hadronic flare appears similar to those obtained in the one-zone leptohadronic models for \hsp~investigated by \cite{2020ApJ...899..113P}, there are three main differences. First, the X-rays of the hadronic flare are the result of proton synchrotron radiation, with some contribution from synchrotron radiation of Bethe-Heitler pairs at energies $\gtrsim10$~keV; in the one-zone leptohadronic models, the X-rays are attributed to synchrotron radiation of accelerated (primary) electrons. Second, the SSC emission, in all hadronic flares we explored, is strongly suppressed due to the strong magnetic field and the ultra-relativistic energies of secondary pairs that make Klein-Nishina effects relevant; in leptohadronic models, the SSC emission may have a significant contribution to the GeV spectrum depending on the parameters. Third, the accompanying neutrino emission of the hadronic flare (at least for the saturated flares studied here) has a comparable (or even higher by a factor of a few) flux to that of the X-ray flare; in the leptohadronic models, however, the neutrino flux was found to be less than the X-ray flare flux by at least a factor of $\sim 3$. 

We next discuss the neutrino expectation of the hadronic flare in the context of the IceCube-200107A detection. We 
compute the number of muon neutrinos and anti-neutrinos, $\mathcal{N}_{\nu_{\mu}+\bar{\nu}_\mu}$ from the source as 
\eqb 
\mathcal{N}_{\nu_{\mu}+\bar{\nu}_\mu} = \frac{1}{3} 
\int_{T_{-}}^{T_{+}} \!\!\! {\rm d}t \int_{\varepsilon_{\nu, \min}}^{\varepsilon_{\nu, \max}} \!\!\!\!\!\!\! {\rm d} \varepsilon_{\nu} \,  A_{\rm eff}(\varepsilon_{\nu},\delta) \phi_{\varepsilon_{\nu}}(t).
\label{eq:Nnu}
\eqe 
Here, $\phi_{\varepsilon_{\nu}}(t)$ is the time-varying all-flavor neutrino and anti-neutrino flux, $T_{\pm}=t_{\nu,pk}\pm\Delta T_{1/2}/2$, $t_{\nu,pk}$ is the peak time of the flare (in the observer's frame), $\Delta T_{1/2}$ is the full width at half maximum of the neutrino flare, $\varepsilon_{\nu, \min}=10^2$~TeV and $\varepsilon_{\nu,\max}= 5\times10^5$~TeV are respectively the minimum and maximum energies considered for the calculation. We also assumed vacuum neutrino mixing and use $1/3$ to convert from the all-flavor to muon neutrino flux. $A_{\rm eff}(\varepsilon_{\nu_{\mu}}, \delta)$ is the energy-dependent and declination-dependent effective area of IceCube. For the latter, we used the IceCube Alert (GFU-All) neutrino effective area\footnote{This is averaged in the declination range $[30^{\circ}-90^{\circ}]$.} from \citet{Blaufuss2019ICRC} and the IceCube Point Source (PS) effective area~\citep{Aartsen:2018ywr}\footnote{Available online at~\url{https://icecube.wisc.edu/science/data}}.  

Table~\ref{tab:nu} summarizes our findings for $\mathcal{N}_{\nu_{\mu}+\bar{\nu}_\mu}$ and $\Delta T_{1/2}$ for all hadronic flares studied so far. To facilitate the comparison between cases, we also list the average neutrino rate, defined as $\langle\dot{\mathcal{N}}_{\nu_{\mu}+\bar{\nu}_{\mu}}\rangle\equiv \mathcal{N}_{\nu_{\mu}+\bar{\nu}_{\mu}}/\Delta T_{1/2}$. The differences in the expected neutrino number among flares are a result of differences in $\Delta T_{1/2}$ and peak neutrino energy. For example, Flares 1 and 3 have similar $\Delta T_{1/2}$, but the latter yields a much higher number of events, because the neutrino spectrum, $\phi_{\varepsilon_\nu}$, peaks closer to the energy where the effective area maximizes. On the contrary, Flares 3 and 5 have different durations, but similar average neutrino rates, suggesting that both flares yield similar neutrino spectra in terms of peak energy and peak flux. 

Flare 5, which we tentatively compared to the recent X-ray flare of \hsp, yields $\mathcal{N}_{\nu_{\mu}+\bar{\nu}_\mu}(>100~\rm TeV)\sim 4\times10^{-4}$ ($\sim2.7\times10^{-3}$)  with the Alert (PS) effective area.  The Poisson probability to detect one or more muon (or anti-muon) neutrinos above 100 TeV within 3 days of the X-ray flare is  $\mathcal{P}|_{1\,\nu_{\mu}\,{\rm or}\,\bar{\nu}_{\mu}} (>100~\rm TeV) \sim 0.04\%$ ($\sim0.27\%$) in the Alert (PS) channels. Similar estimates have been presented by \cite{2020ApJ...899..113P} in the context of standard leptohadronic models in which the X-ray flare is produced by the synchrotron radiation of accelerated electrons, and the $\gamma$-ray flux has contributions from secondary pairs. For comparison, their most optimistic leptohadronic model (see Model $A_{B'=30~\rm G}$ in Table 3 of \citealt{2020ApJ...899..113P}) predicts $\mathcal{N}_{\nu_{\mu}+\bar{\nu}_\mu}(>100~\rm TeV)\sim 0.5\times10^{-4}$ and $5.9\times10^{-4}$ in 4 days in the Alert and PS effective areas, respectively.  These numbers are lower by a factor of $\sim 8$ (4.5) when compared to the expectation of Flare 5 for the Alert (PS) effective area. 
\emph{Thus, the neutrino number expected during saturated hadronic X-ray flares may be higher  than the one predicted in standard leptohadronic models by a factor of $\sim3-10$ depending on source parameters.}

Based on our neutrino estimates, it is  unlikely that a single hadronic X-ray flare, such as Flare 5 (or any other flare from those studied here for that matter), can produce a neutrino event, like \icv. For example, $N_f \approx 63$ flares with the same exact properties as Flare 5 would yield $\mathcal{N}_{\nu_{\mu}+\bar{\nu}_\mu} \sim 0.17$ in an IceCube point source search of \hsp. This expectation would correspond to the $84\%$ Poisson lower limit for the detection of 1 event. However, the $\sim 7$~year-long gap in the X-ray light curve of \hsp~makes it difficult to test this scenario. We plan to test the implications of hadronic saturated X-ray flares in a future work by estimating the number of muon neutrinos and anti-neutrinos expected from other extreme blazars that were observed by \swift~more than 100 times and were detected in a flaring state (Giommi et al., 2020, in prep.).

\begin{deluxetable}{cccc}
\centering
\tablecaption{Full width at half maximum of neutrino flares, number of muon and antimuon neutrinos with energy $>100$~TeV expected to be detected by IceCube with the Alert (Point Source) searches within that duration, and average neutrino rate, defined as $\langle\dot{\mathcal{N}}_{\nu_{\mu}+\bar{\nu}_{\mu}}\rangle\equiv \mathcal{N}_{\nu_{\mu}+\bar{\nu}_{\mu}}/\Delta T_{1/2}$. \label{tab:nu}}
\tablewidth{0pt}
 \tablehead{
 \colhead{Flare} & \colhead{$\Delta T_{1/2}$ (d)}  & $\mathcal{N}_{\nu_{\mu}+\bar{\nu}_{\mu}}$ ($\times 10^{-4}$) & $\langle\dot{\mathcal{N}}_{\nu_{\mu}+\bar{\nu}_{\mu}}\rangle$ ($\times 10^{-4}$ d$^{-1}$)\\
 &  & Alert (Point Source)  &  Alert (Point Source)
 }
 \startdata
 1  & 0.96 &  0.003 (0.03) & 0.003 (0.03) \\
 2 & 0.42 & 0.11 (0.78) & 0.25 (1.87) \\ 
 3 & 0.66 &  0.85 (5.18) & 1.29 (7.83) \\
 4 & 2.04 &  1.79 (9.00) & 0.87 (4.41) \\ 
 5 & 3.05 & 4.00 (27.3) & 1.31 (8.97) \\
\enddata
\tablecomments{The estimates were made for a blazar at the redshift and declination of \hsp.}
\end{deluxetable}

\section{Prospects for future observatories}\label{sec:future}
We have shown that saturated hadronic flares at keV energies are accompanied by an equally bright MeV flare.  This soft $\gamma$-ray component has a broad curved energy spectrum (see Figure~\ref{fig2}) and originates from the synchrotron radiation of secondary pairs from the Bethe-Heitler process \citep[see also][]{2015MNRAS.447...36P}. As a result, it is a unique electromagnetic signature of the proposed hadronic saturated flares. The predicted synchrotron MeV flares are ideal signals to be searched for with future sensitive MeV observatories with wide field of view, good spectral resolution, and polarization capabilities, such as  the All-sky Medium Energy Gamma-ray Observatory (\emph{AMEGO})~\citep{2019BAAS...51g.245M} and \emph{e-ASTROGAM}~\citep{2017ExA....44...25D}. 

\begin{figure}
\centering
\includegraphics[width=0.49\textwidth]{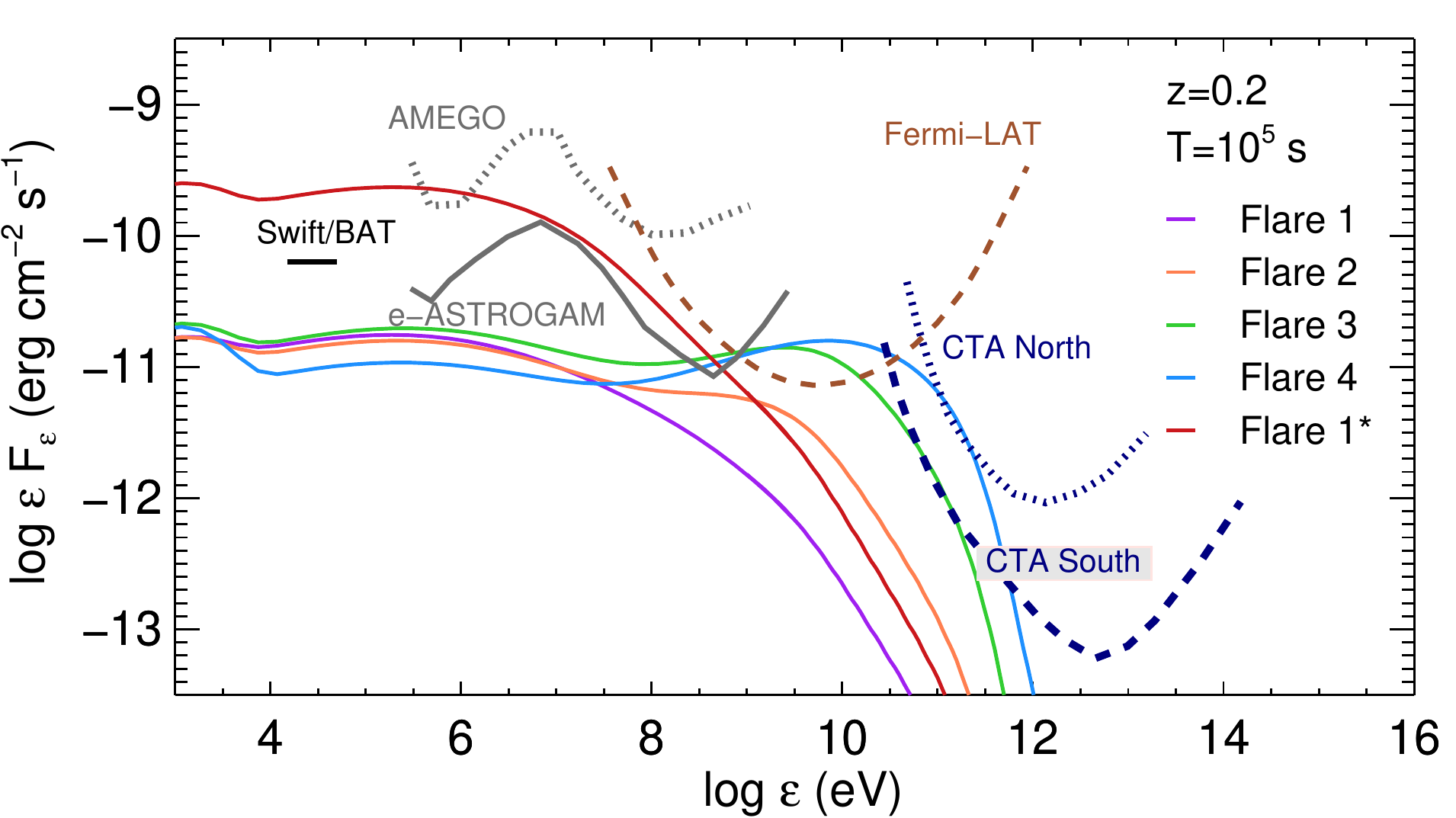}
\caption{$\gamma$-ray energy spectra (solid colored lines) of a fiducial source at redshift $z=0.2$ exhibiting hadronic X-ray flares for the same parameters as in Figure~\ref{fig2}. Results for a flare with 15 times higher peak X-ray luminosity than Flare~1 are also shown in dark red. Spectra are averaged over a time-window of duration $10^5$~s, centered around the peak time of each flare. All $\gamma$-ray spectra have been attenuated using the EBL model of \citet{2010ApJ...712..238F}. The black horizontal line indicates the median $1\sigma$ sensitivity of \swift/BAT for a temporal exposure of 1~day \citep{2013ApJS..209...14K}. Other lines show the sensitivity of current and future $\gamma$-ray instruments, all scaled to an exposure time of $10^5$~s: \emph{Fermi}-LAT (adopted from \url{https://www.slac.stanford.edu/exp/glast/groups/canda/lat_Performance.htm} for a high galactic latitude), \emph{e-ASTROGAM}~\citep{2017ExA....44...25D}, \emph{AMEGO}~\citep{2019BAAS...51g.245M}, and CTA (adopted from \citealt{2020JPhCS1468a2096H}). }
    \label{fig5}
\end{figure}

In Figure~\ref{fig5} we show the $\gamma$-ray spectra (solid colored lines) for the four hadronic flares discussed in Section~\ref{sec:results} that reach peak luminosities of  $\sim 10^{45}$~erg s$^{-1}$. An additional case with 15 times higher peak X-ray luminosity than Flare 1 is also included in the plot (indicated as Flare 1*). The source is assumed to be located at $z=0.2$. All spectra are averaged over a period of $10^5$~s, thus resulting in average fluxes that are 2-3 times lower than the corresponding peak flare fluxes. The $\gamma$-ray spectra have been attenuated due to the extragalactic background light (EBL) using the model of \cite{2010ApJ...712..238F}. The sensitivity curves of existing and future $\gamma$-ray instruments, covering a wide energy range ($\sim 1$~MeV--$100$~TeV), are overplotted for comparison. All curves are scaled to the same exposure time, $t_{\rm exp}=10^5$~s, assuming that the limiting flux is $\propto 1/\sqrt{t_{\rm exp}}$. For comparison, we also show the median $1\sigma$ sensitivity of the \swift \, Burst Alert Telescope (BAT) for a temporal exposure of 1~day \citep{2013ApJS..209...14K}.

X-ray flares from less compact emitting regions with weaker magnetic fields are accompanied by sub-TeV emission (see e.g., Flares 3 and 4) that could be detected by the Cherenkov Telescope Array  \citep[CTA,][]{2017APh....93...76H}. In this regard, nearby HSP blazars at bright X-ray flaring states are ideal targets for the search of hadronic flares with CTA. Meanwhile, not all hadronic flares are strong $\gamma$-ray emitters in the 10--100~GeV energy range, wherein the sensitivity of \fermi-LAT is best. Flare 1* (solid dark red line) is an indicative example of hadronic flares that are hidden in GeV $\gamma$-rays, but are still strong neutrino emitters. In this particular example, the peak neutrino luminosity, which is comparable to the peak  luminosity of the flare at keV energies, is $\sim 8\times10^{46}$~erg s$^{-1}$. The attenuated $\gamma$-ray emission of such flares is not lost but re-emerges at lower photon energies as a broad MeV component with luminosity comparable to the X-ray and neutrino luminosities.  Such bright flares could be detected in hard X-rays (i.e., $15-50$~keV) with \swift/BAT and would lie within the detection capabilities of proposed X-ray and  MeV $\gamma$-ray satellites, like \emph{STROBE-X}\footnote{The Wide-Field Monitor, with a designed 1-day sensitivity of $\sim 2$~mCrab in the $2-50$~keV energy range, will be ideal for blazar X-ray monitoring and triggering of pointed observations of bright X-ray flares.} \citep{2019arXiv190303035R} and \emph{e-ASTROGAM}, respectively. Therefore, our results motivate future searches for neutrino counterparts to blazar flares in hard X-rays and soft $\gamma$-rays. 

Another common property of hadronic X-ray flares  is that their broadband spectrum (i.e., from optical up to soft $\gamma$-rays) is produced by the synchrotron radiation of relativistic protons and their secondary pairs. The contribution of synchrotron self-Compton radiation of the latter in the optical to soft-$\gamma$-ray flux is negligible for the cases explored here. As a result, the polarization degree of a hadronic X-ray flare is expected to be almost constant across a wide range of frequencies, with the actual value depending on the degree of order (or disorder) of the magnetic field in the emission region \citep{2013ApJ...774...18Z, 2018ApJ...863...98P, 2019ApJ...876..109Z}. The polarization properties of such hadronic flares are expected to differ from  those characterizing the average flux state of a blazar, as their production region may be other than the typical blazar radiation zone. X-ray polarization of bright blazar flares  can be measured in the near future \citep[see e.g.,][]{2019ApJ...880...29L}  by the Imaging X-ray Polarimetry Explorer \citep[\emph{IXPE},][]{IXPE2018}, thus providing some hints to their hadronic or leptonic origin. In the long term, combined polarimetric observations of blazars in flaring states in the optical, X-ray and soft $\gamma$-ray energy ranges can help to test the scenario of purely hadronic X-ray flares.

\section{Discussion}\label{sec:discussion}

In this paper we studied hadronic saturated flares in the case where the source parameters are such as to produce X-rays from proton synchrotron radiation. While hadronic saturated flares peaking at lower energies (e.g., in the optical) are also possible, extremely high proton energies would be required to meet the threshold condition for photopion production on the proton synchrotron photons, namely $\gamma'_p \gtrsim 1.5\times10^9 \varepsilon^{-1}_{ph, \rm eV}\mathcal{D}_1$ (where we used equations \ref{eq:xpsyn} and \ref{eq:pg-thr-psyn}, and $\varepsilon_{ph, \rm eV}\equiv \varepsilon_{ph}/1~{\rm eV}$). The magnetic field necessary to power these fiducial optical flares with an accompanying neutrino flare would be $B' \lesssim 7\, \varepsilon^{3}_{ph, \rm eV} \mathcal{D}^{-3}_1$~nG. Such extremely low  values are unrealistic for the parsec (pc) scale AGN jets, while leading to Mpc-scale proton gyroradii.  Meanwhile, the requirements for the photomeson threshold condition on proton synchrotron photons are easily met for hadronic flares peaking at MeV energies. However, the associated neutrino flux would be lower than that found in hadronic X-ray flares, because the density of target photons (from proton synchrotron radiation) would be significantly reduced. The ratio of the photon number densities at X-ray and MeV energies can be written as $n_{\rm keV}/n_{\rm MeV}\approx \left(\varepsilon_{ph, \rm keV}/\varepsilon_{ph, \rm MeV}\right)^{-1} \ll 1$, where we assumed that both flares have the same peak luminosity and approximated the proton synchrotron spectrum as monoenergetic. Only the presence of an ambient dense radiation field (peaking in X-rays in the jet frame) would increase the neutrino production efficiency of hadronic saturated MeV flares, but this requires a separate study.

We found that saturated hadronic X-ray flares exhibit some interesting phenomenology (see Section~\ref{sec:results}). One of our main findings is that the energy flux of a saturated hadronic X-ray flares is a good proxy for the bolometric all-flavor neutrino energy flux. Similar conclusions were also drawn for the 2017 multi-wavelength flare of \txs, even though the emission model for the electromagnetic emission was different \citep[e.g.,][]{Keivani2018, 2018ApJ...865..124M, Gao2019}. More specifically, different studies showed that the SED of the 2017 flare \citep[see also][for multi-epoch observations of \txs]{2020ApJ...891..115P} can be well explained by synchrotron and inverse Compton processes of accelerated electrons in the blazar jet. As a result, any emission arising from proton-initiated cascades in the source should not exceed the observed X-ray flux. Given that in ISP blazars like \txs, X-rays fall in the trough between the two SED humps, the hadronic component, responsible for high-energy neutrino production, was found to be  radiatively subdominant, and  $L_{\nu} \lesssim L_{X}$. A correlation between the X-ray and high-energy neutrino luminosity is also expected in leptohadronic emission models of HSPs \citep[e.g.,][]{2015MNRAS.447...36P, 2015MNRAS.448..910C}. According to these scenarios, there is a significant contribution of secondaries from photohadronic interactions to the observed $\gamma$-ray flux, while the low-peak energy photons that are produced by electron synchrotron radiation (X-rays for the case of HSPs) are the main target photons for neutrino production.  Typically, the leptohadronic emission models, especially for BL Lacs, are less efficient in neutrino production than the saturated hadronic flares studied here. As a result, $L_{\nu}< L_X$ \citep[see also][for application to an extreme HSP]{2020ApJ...899..113P}. When leptohadronic models are applied to the SEDs of flat spectrum radio quasars (FSRQs), the level of expected neutrino flux is not always bounded by the X-ray flux. Usually, the proton-initiated cascade emission peaks in $\gamma$-rays, and the X-ray band probes the low-energy tail of the cascade spectrum. Thus, $L_{\nu} > L_X$ and particularly during flaring $\gamma$-ray states \citep[see e.g.,][]{2017ApJ...843..109G, 2020arXiv200904026R}.

We also found that the (sub-)MeV emission of a saturated hadronic flare is a second proxy for high-energy neutrino emission. Close to the peak time of the X-ray flare, an equally bright broad spectral component peaks in the MeV energy range. This component is a result of the electromagnetic cascades developed in the source from secondary pairs and, more specifically, from those produced via $\gamma \gamma$ absorption \citep[see also][]{2015MNRAS.447...36P, 2019ApJ...874L..29R, 2019ApJ...876..109Z}. This accompanying MeV emission  
might provide a complementary probe of hadronic flares from nearby sources with the advent of missions like \emph{AMEGO} and \emph{e-ASTROGAM}.  On the contrary, flaring  in GeV-TeV $\gamma$-rays
is not always expected as it depends on the source parameters (see Figure~\ref{fig3}).  More specifically, for compact emitting regions with strong magnetic fields, the source becomes opaque in $\gamma$-rays  and the high-energy neutrino flare may not have a bright $\gamma$-ray electromagnetic counterpart. Thus, neutrino-rich hadronic X-ray flares may be dark in GeV/TeV $\gamma$-rays. This finding is consistent with the conclusions of previous studies about the origin of the IceCube diffuse neutrino flux \citep[see e.g.,][]{2016PhRvL.116g1101M, 2020arXiv200109520I} and the neutrino flare of \txs~in 2014/15 \citep[][]{2018ApJ...865..124M, 2019ApJ...881...46R,2019ApJ...874L..29R,  2020ApJ...891..115P, 2020ApJ...889..118Z}.  

We applied the idea of saturated hadronic flares to to the blazar \hsp~(see Figure~\ref{fig4}), which has been possibly associated with a high-energy neutrino (IceCube-200107A) while undergoing a hard X-ray flare. We found that the number of neutrinos above 100~TeV during saturated hadronic flares can be up to $3-10$ times higher than the expected number in standard leptohadronic models \citep{2020ApJ...899..113P}. Still, it 
is unlikely that a single X-ray flare can explain the detected neutrino event \icv. In the best case scenario, we found that approximately 63 (3-day long) X-ray flares would yield $\sim 0.17$ muon and antimuon neutrinos above 100 TeV in an IceCube point source search of \hsp. This translates into a $\sim 5\%$ duty cycle of hadronic X-ray flaring activity, considering a total period of 10 years. The identification of X-ray counterparts to high-energy neutrinos  in the post-\swift~era  and the establishment of the X-ray flaring duty cycle in blazars with future instruments like \emph{STROBE-X} \citep{2019arXiv190303035R} are essential for testing the proposed theoretical scenario and narrowing down the candidate neutrino sources. 

Although we applied the scenario of hadronic X-ray flares to the extreme blazar \hsp, the proposed model can be still of relevance to intermediate- and low-synchrotron peaked blazars  (ISPs and LSPs, respectively). 
For example, in LSPs where the bolometric radiation is emitted as GeV $\gamma$-rays, hadronic X-ray flares would make only a small contribution to the bolometric luminosity of the source. Even in this case, however, the prediction of an accompanying TeV--PeV neutrino flare with comparable luminosity to the X-ray one would still hold. An application of our proposed model to X-ray flares detected by the \swift \, X-Ray Telescope (XRT) from different blazar subclasses will be the subject of a forthcoming publication.

In this paper we focused on purely hadronic flares by neglecting the injection of relativistic primary electrons in the flaring region. This could be physically realized in the following ways. One idea is that  both primary electrons and protons after being accelerated to relativistic energies, escape to a region where strong radiative losses drain the energy from the electrons but do not strongly limit the energy of protons. Being still energetic, protons can enter another region producing a hadronic flare, while the  low-energy primary electrons making a negligible contribution to the photon emission. In this scenario, a leptonic flare powered by primary electrons will be produced in the first radiation zone, while a hadronic flare will arise from the second zone. The spectral and temporal properties of such multi-zone flaring model require dedicated investigation. Alternatively, primary relativistic electrons can be injected directly to the production region of the hadronic flare with low enough luminosity so that they are radiatively sub-dominant. 
This can be exemplified in the case of Flare 5 that is constrained by observations. Because  proton synchrotron radiation is responsible for the X-rays, any primary electron contribution should remain below the observed flux. This implies that $L_e<(m_e/m_p)L_p$ where $L_e$ is the primary electron luminosity. 

 In our analysis of saturated hadronic flares we did not include any external radiation fields that may be present in the AGN environment (from e.g., the broad line region or the dusty torus). This implicit assumption is valid as long as the number density of external photons is lower than the number density of proton synchrotron photons and/or whenever the collisions between external photons and relativistic protons are below the threshold for photomeson production.  Inclusion of non-negligible external photon fields with typical energies in the optical/UV or near-infrared bands would still not affect the development of the hadronic flares in any way other than shifting the peak energy of the neutrino spectrum. 
 
\section{Conclusion}\label{sec:conclusion}
Motivated by the tentative associations of IceCube high-energy neutrinos with blazars and the lack of $\gamma$-ray flaring activity that is often found close to the neutrino arrival time, we investigated in a fully time-dependent manner, neutrino production in coincidence with X-ray flares that are powered by proton synchrotron radiation. Such hadronic flares may occur whenever protons are accelerated intermittently to high enough energies in the jet, and produce pions interacting mainly with their own synchrotron radiation. 

The proposed hadronic X-ray flares pose an interesting alternative to the standard leptohadronic scenarios for neutrino production in blazars. They are accompanied by equally bright TeV--PeV neutrino flares and MeV $\gamma$-ray flares, while they can be hidden in GeV--TeV $\gamma$-rays for dense source environments. Our results motivate X-ray and MeV $\gamma$-ray monitoring of blazars for the search of high-energy neutrino counterparts.  


\acknowledgments
The authors would like to thank the anonymous referee for a constructive report. The authors would also like to thank Paolo Padovani, Kohta Murase, Yannis Liodakis and Sara Buson for their useful comments on the manuscript. A.M. would like to thank the Department of Astrophysical Sciences at Princeton University for its hospitality during which this project was conceived. M.P. acknowledges support from the Lyman Jr.~Spitzer Postdoctoral Fellowship and NASA Fermi grant No.~80NSSC18K1745.

\bibliography{blazar}

\begin{thebibliography}{}
\expandafter\ifx\csname natexlab\endcsname\relax\def\natexlab#1{#1}\fi
\providecommand{\url}[1]{\href{#1}{#1}}
\providecommand{\dodoi}[1]{doi:~\href{http://doi.org/#1}{\nolinkurl{#1}}}
\providecommand{\doeprint}[1]{\href{http://ascl.net/#1}{\nolinkurl{http://ascl.net/#1}}}
\providecommand{\doarXiv}[1]{\href{https://arxiv.org/abs/#1}{\nolinkurl{https://arxiv.org/abs/#1}}}

\bibitem[{Aartsen {et~al.}(2013{\natexlab{a}})}]{Aartsen:2013a}
Aartsen, M., {et~al.} 2013{\natexlab{a}}, Phys.Rev.Lett., 111, 021103,
  \dodoi{10.1103/PhysRevLett.111.021103}

\bibitem[{Aartsen {et~al.}(2013{\natexlab{b}})}]{Aartsen:2013b}
---. 2013{\natexlab{b}}, Science, 342, 1242856, \dodoi{10.1126/science.1242856}

\bibitem[{{Abdollahi} {et~al.}(2020){Abdollahi}, {Acero}, {Ackermann},
  {Ajello}, {Atwood}, {Axelsson}, {Baldini}, {Ballet}, {Barbiellini},
  {Bastieri}, {Becerra Gonzalez}, {Bellazzini}, {Berretta}, {Bissaldi}, {Bland
  ford}, {Bloom}, {Bonino}, {Bottacini}, {Brandt}, {Bregeon}, {Bruel},
  {Buehler}, {Burnett}, {Buson}, {Cameron}, {Caputo}, {Caraveo}, {Casandjian},
  {Castro}, {Cavazzuti}, {Charles}, {Chaty}, {Chen}, {Cheung}, {Chiaro},
  {Ciprini}, {Cohen-Tanugi}, {Cominsky}, {Coronado-Bl{\'a}zquez}, {Costantin},
  {Cuoco}, {Cutini}, {D'Ammando}, {DeKlotz}, {Torre Luque}, {de Palma},
  {Desai}, {Digel}, {Lalla}, {Mauro}, {Venere}, {Dom{\'\i}nguez}, {Dumora},
  {Dirirsa}, {Fegan}, {Ferrara}, {Franckowiak}, {Fukazawa}, {Funk}, {Fusco},
  {Gargano}, {Gasparrini}, {Giglietto}, {Giommi}, {Giordano}, {Giroletti},
  {Glanzman}, {Green}, {Grenier}, {Griffin}, {Grondin}, {Grove}, {Guiriec},
  {Harding}, {Hayashi}, {Hays}, {Hewitt}, {Horan}, {J{\'o}hannesson},
  {Johnson}, {Kamae}, {Kerr}, {Kocevski}, {Kovac'evic'}, {Kuss}, {Landriu},
  {Larsson}, {Latronico}, {Lemoine-Goumard}, {Li}, {Liodakis}, {Longo},
  {Loparco}, {Lott}, {Lovellette}, {Lubrano}, {Madejski}, {Maldera},
  {Malyshev}, {Manfreda}, {Marchesini}, {Marcotulli}, {Mart{\'\i}-Devesa},
  {Martin}, {Massaro}, {Mazziotta}, {McEnery}, {Mereu}, {Meyer}, {Michelson},
  {Mirabal}, {Mizuno}, {Monzani}, {Morselli}, {Moskalenko}, {Negro}, {Nuss},
  {Ojha}, {Omodei}, {Orienti}, {Orlando}, {Ormes}, {Palatiello}, {Paliya},
  {Paneque}, {Pei}, {Pe{\~n}a-Herazo}, {Perkins}, {Persic}, {Pesce-Rollins},
  {Petrosian}, {Petrov}, {Piron}, {Poon}, {Porter}, {Principe}, {Rain{\`o}},
  {Rando}, {Razzano}, {Razzaque}, {Reimer}, {Reimer}, {Remy}, {Reposeur},
  {Romani}, {Parkinson}, {Schinzel}, {Serini}, {Sgr{\`o}}, {Siskind}, {Smith},
  {Spandre}, {Spinelli}, {Strong}, {Suson}, {Tajima}, {Takahashi}, {Tak},
  {Thayer}, {Thompson}, {Tibaldo}, {Torres}, {Torresi}, {Valverde}, {Klaveren},
  {Zyl}, {Wood}, {Yassine}, \& {Zaharijas}}]{2020ApJS..247...33A}
{Abdollahi}, S., {Acero}, F., {Ackermann}, M., {et~al.} 2020, \apjs, 247, 33,
  \dodoi{10.3847/1538-4365/ab6bcb}

\bibitem[{{Aharonian}(2000)}]{2000NewA....5..377A}
{Aharonian}, F.~A. 2000, \na, 5, 377, \dodoi{10.1016/S1384-1076(00)00039-7}

\bibitem[{{Bednarek} \& {Protheroe}(1999)}]{1999MNRAS.302..373B}
{Bednarek}, W., \& {Protheroe}, R.~J. 1999, \mnras, 302, 373,
  \dodoi{10.1046/j.1365-8711.1999.02132.x}

\bibitem[{{Begelman} {et~al.}(1990){Begelman}, {Rudak}, \&
  {Sikora}}]{1990ApJ...362...38B}
{Begelman}, M.~C., {Rudak}, B., \& {Sikora}, M. 1990, \apj, 362, 38,
  \dodoi{10.1086/169241}

\bibitem[{{Biteau} {et~al.}(2020){Biteau}, {Prandini}, {Costamante}, {Lemoine},
  {Padovani}, {Pueschel}, {Resconi}, {Tavecchio}, {Taylor}, \&
  {Zech}}]{2020NatAs...4..124B}
{Biteau}, J., {Prandini}, E., {Costamante}, L., {et~al.} 2020, Nature
  Astronomy, 4, 124, \dodoi{10.1038/s41550-019-0988-4}

\bibitem[{Blaufuss {et~al.}(2020)Blaufuss, Kintscher, Lu, \&
  Tung}]{Blaufuss2019ICRC}
Blaufuss, E., Kintscher, T., Lu, L., \& Tung, C.~F. 2020, PoS, ICRC2019, 1021,
  \dodoi{10.22323/1.358.1021}

\bibitem[{{Cerruti} {et~al.}(2019){Cerruti}, {Zech}, {Boisson}, {Emery},
  {Inoue}, \& {Lenain}}]{2019MNRAS.483L..12C}
{Cerruti}, M., {Zech}, A., {Boisson}, C., {et~al.} 2019, \mnras, 483, L12,
  \dodoi{10.1093/mnrasl/sly210}

\bibitem[{{Cerruti} {et~al.}(2015){Cerruti}, {Zech}, {Boisson}, \&
  {Inoue}}]{2015MNRAS.448..910C}
{Cerruti}, M., {Zech}, A., {Boisson}, C., \& {Inoue}, S. 2015, \mnras, 448,
  910, \dodoi{10.1093/mnras/stu2691}

\bibitem[{{Chang} {et~al.}(2019){Chang}, {Arsioli}, {Giommi}, {Padovani}, \&
  {Brandt}}]{2019A&A...632A..77C}
{Chang}, Y.~L., {Arsioli}, B., {Giommi}, P., {Padovani}, P., \& {Brandt}, C.~H.
  2019, \aap, 632, A77, \dodoi{10.1051/0004-6361/201834526}

\bibitem[{{De Angelis} {et~al.}(2017){De Angelis}, {Tatischeff}, {Tavani},
  {Oberlack}, {Grenier}, {Hanlon}, {Walter}, {Argan}, {von Ballmoos},
  {Bulgarelli}, {Donnarumma}, {Hernanz}, {Kuvvetli}, {Pearce}, {Zdziarski},
  {Aboudan}, {Ajello}, {Ambrosi}, {Bernard}, {Bernardini}, {Bonvicini},
  {Brogna}, {Branchesi}, {Budtz-Jorgensen}, {Bykov}, {Campana}, {Cardillo},
  {Coppi}, {De Martino}, {Diehl}, {Doro}, {Fioretti}, {Funk}, {Ghisellini},
  {Grove}, {Hamadache}, {Hartmann}, {Hayashida}, {Isern}, {Kanbach}, {Kiener},
  {Kn{\"o}dlseder}, {Labanti}, {Laurent}, {Limousin}, {Longo}, {Mannheim},
  {Marisaldi}, {Martinez}, {Mazziotta}, {McEnery}, {Mereghetti}, {Minervini},
  {Moiseev}, {Morselli}, {Nakazawa}, {Orleanski}, {Paredes}, {Patricelli},
  {Peyr{\'e}}, {Piano}, {Pohl}, {Ramarijaona}, {Rando}, {Reichardt},
  {Roncadelli}, {Silva}, {Tavecchio}, {Thompson}, {Turolla}, {Ulyanov},
  {Vacchi}, {Wu}, \& {Zoglauer}}]{2017ExA....44...25D}
{De Angelis}, A., {Tatischeff}, V., {Tavani}, M., {et~al.} 2017, Experimental
  Astronomy, 44, 25, \dodoi{10.1007/s10686-017-9533-6}

\bibitem[{{Dimitrakoudis} {et~al.}(2012){Dimitrakoudis}, {Mastichiadis},
  {Protheroe}, \& {Reimer}}]{2012A&A...546A.120D}
{Dimitrakoudis}, S., {Mastichiadis}, A., {Protheroe}, R.~J., \& {Reimer}, A.
  2012, \aap, 546, A120, \dodoi{10.1051/0004-6361/201219770}

\bibitem[{{Finke} {et~al.}(2010){Finke}, {Razzaque}, \&
  {Dermer}}]{2010ApJ...712..238F}
{Finke}, J.~D., {Razzaque}, S., \& {Dermer}, C.~D. 2010, \apj, 712, 238,
  \dodoi{10.1088/0004-637X/712/1/238}

\bibitem[{{Gao} {et~al.}(2019){Gao}, {Fedynitch}, {Winter}, \&
  {Pohl}}]{Gao2019}
{Gao}, S., {Fedynitch}, A., {Winter}, W., \& {Pohl}, M. 2019, Nature Astronomy,
  3, 88, \dodoi{10.1038/s41550-018-0610-1}

\bibitem[{{Gao} {et~al.}(2017){Gao}, {Pohl}, \& {Winter}}]{2017ApJ...843..109G}
{Gao}, S., {Pohl}, M., \& {Winter}, W. 2017, \apj, 843, 109,
  \dodoi{10.3847/1538-4357/aa7754}

\bibitem[{{Garrappa} {et~al.}(2020){Garrappa}, {Buson}, \& {Fermi-LAT
  Collaboration}}]{2020GCN.26669....1G}
{Garrappa}, S., {Buson}, S., \& {Fermi-LAT Collaboration}. 2020, GRB
  Coordinates Network, 26669, 1

\bibitem[{{Giommi} {et~al.}(2020{\natexlab{a}}){Giommi}, {Glauch}, \&
  {Resconi}}]{2020ATel13394....1G}
{Giommi}, P., {Glauch}, T., \& {Resconi}, E. 2020{\natexlab{a}}, The
  Astronomer's Telegram, 13394, 1

\bibitem[{{Giommi} {et~al.}(2020{\natexlab{b}}){Giommi}, {Padovani},
  {Oikonomou}, {Glauch}, {Paiano}, \& {Resconi}}]{2020A&A...640L...4G}
{Giommi}, P., {Padovani}, P., {Oikonomou}, F., {et~al.} 2020{\natexlab{b}},
  \aap, 640, L4, \dodoi{10.1051/0004-6361/202038423}

\bibitem[{{Hassan} {et~al.}(2017){Hassan}, {Arrabito}, {Bernl{\"o}hr},
  {Bregeon}, {Cortina}, {Cumani}, {Di Pierro}, {Falceta-Goncalves}, {Lang},
  {Hinton}, {Jogler}, {Maier}, {Moralejo}, {Morselli}, {Todero Peixoto}, \&
  {Wood}}]{2017APh....93...76H}
{Hassan}, T., {Arrabito}, L., {Bernl{\"o}hr}, K., {et~al.} 2017, Astroparticle
  Physics, 93, 76, \dodoi{10.1016/j.astropartphys.2017.05.001}

\bibitem[{{Hinton} \& {Ruiz-Velasco}(2020)}]{2020JPhCS1468a2096H}
{Hinton}, J., \& {Ruiz-Velasco}, E. 2020, in Journal of Physics Conference
  Series, Vol. 1468, Journal of Physics Conference Series, 012096,
  \dodoi{10.1088/1742-6596/1468/1/012096}

\bibitem[{{IceCube Collaboration}(2018)}]{IceCube:2018cha}
{IceCube Collaboration}. 2018, Science, 361, 147,
  \dodoi{10.1126/science.aat2890}

\bibitem[{{IceCube Collaboration}(2019)}]{Aartsen:2018ywr}
---. 2019, Eur. Phys. J., C79, 234, \dodoi{10.1140/epjc/s10052-019-6680-0}

\bibitem[{{IceCube Collaboration}(2020)}]{2020GCN.26655....1I}
---. 2020, GRB Coordinates Network, 26655, 1

\bibitem[{{IceCube Collaboration} {et~al.}(2018){IceCube Collaboration},
  Fermi-LAT, MAGIC, AGILE, ASAS-SN, HAWC, H.E.S.S., INTEGRAL, Kanata, Kiso,
  Kapteyn, Telescope, Subaru, Swift, NuSTAR, VERITAS, \&
  teams}]{IceCube:2018dnn}
{IceCube Collaboration}, Fermi-LAT, MAGIC, {et~al.} 2018, Science, 361,
  eaat1378, \dodoi{10.1126/science.aat1378}

\bibitem[{{IceCube Collaboration} {et~al.}(2020){IceCube Collaboration},
  {Aartsen}, {Ackermann}, {Adams}, {Aguilar}, {Ahlers}, {Ahrens}, {Alispach},
  {Andeen}, {Anderson}, {Ansseau}, {Anton}, {Arg{\"u}elles}, {Auffenberg},
  {Axani}, {Backes}, {Bagherpour}, {Bai}, {Balagopal V.}, {Barbano}, {Barwick},
  {Bastian}, {Baum}, {Baur}, {Bay}, {Beatty}, {Becker}, {Becker Tjus},
  {BenZvi}, {Berley}, {Bernardini}, {Besson}, {Binder}, {Bindig}, {Blaufuss},
  {Blot}, {Bohm}, {B{\"o}ser}, {Botner}, {B{\"o}ttcher}, {Bourbeau},
  {Bourbeau}, {Bradascio}, {Braun}, {Bron}, {Brostean-Kaiser}, {Burgman},
  {Buscher}, {Busse}, {Carver}, {Chen}, {Cheung}, {Chirkin}, {Choi}, {Clark},
  {Classen}, {Coleman}, {Collin}, {Conrad}, {Coppin}, {Correa}, {Cowen},
  {Cross}, {Dave}, {De Clercq}, {DeLaunay}, {Dembinski}, {Deoskar}, {De
  Ridder}, {Desiati}, {de Vries}, {de Wasseige}, {de With}, {DeYoung}, {Diaz},
  {D{\'\i}az-V{\'e}lez}, {Dujmovic}, {Dunkman}, {Dvorak}, {Eberhardt},
  {Ehrhardt}, {Eller}, {Engel}, {Evenson}, {Fahey}, {Fazely}, {Felde},
  {Filimonov}, {Finley}, {Fox}, {Franckowiak}, {Friedman}, {Fritz}, {Gaisser},
  {Gallagher}, {Ganster}, {Garrappa}, {Gerhardt}, {Ghorbani}, {Glauch},
  {Gl{\"u}senkamp}, {Goldschmidt}, {Gonzalez}, {Grant}, {Gr{\'e}goire},
  {Griffith}, {Griswold}, {G{\"u}nder}, {G{\"u}nd{\"u}z}, {Haack}, {Hallgren},
  {Halliday}, {Halve}, {Halzen}, {Hanson}, {Haungs}, {Hebecker}, {Heereman},
  {Heix}, {Helbing}, {Hellauer}, {Henningsen}, {Hickford}, {Hignight}, {Hill},
  {Hoffman}, {Hoffmann}, {Hoinka}, {Hokanson-Fasig}, {Hoshina}, {Huang},
  {Huber}, {Huber}, {Hultqvist}, {H{\"u}nnefeld}, {Hussain}, {In}, {Iovine},
  {Ishihara}, {Jansson}, {Japaridze}, {Jeong}, {Jero}, {Jones}, {Jonske},
  {Joppe}, {Kang}, {Kang}, {Kappes}, {Kappesser}, {Karg}, {Karl}, {Karle},
  {Katz}, {Kauer}, {Kelley}, {Kheirandish}, {Kim}, {Kintscher}, {Kiryluk},
  {Kittler}, {Klein}, {Koirala}, {Kolanoski}, {K{\"o}pke}, {Kopper}, {Kopper},
  {Koskinen}, {Kowalski}, {Krings}, {Kr{\"u}ckl}, {Kulacz}, {Kurahashi},
  {Kyriacou}, {Lanfranchi}, {Larson}, {Lauber}, {Lazar}, {Leonard},
  {Lesiak-Bzdak}, {Leszczy{\'n}ska}, {Leuermann}, {Liu}, {Lohfink}, {Lozano
  Mariscal}, {Lu}, {Lucarelli}, {L{\"u}nemann}, {Luszczak}, {Lyu}, {Ma},
  {Madsen}, {Maggi}, {Mahn}, {Makino}, {Mallik}, {Mallot}, {Mancina},
  {Mari\{{\textcommabelow s}\}}, {Maruyama}, {Mase}, {Maunu}, {McNally},
  {Meagher}, {Medici}, {Medina}, {Meier}, {Meighen-Berger}, {Merino}, {Meures},
  {Micallef}, {Mockler}, {Moment{\'e}}, {Montaruli}, {Moore}, {Morse},
  {Moulai}, {Muth}, {Nagai}, {Naumann}, {Neer}, {Niederhausen}, {Nisa},
  {Nowicki}, {Nygren}, {Obertacke Pollmann}, {Oehler}, {Olivas}, {O'Murchadha},
  {O'Sullivan}, {Palczewski}, {Pandya}, {Pankova}, {Park}, {Peiffer},
  {P{\'e}rez de los Heros}, {Philippen}, {Pieloth}, {Pinat}, {Pizzuto}, {Plum},
  {Porcelli}, {Price}, {Przybylski}, {Raab}, {Raissi}, {Rameez}, {Rauch},
  {Rawlins}, {Rea}, {Rehman}, {Reimann}, {Relethford}, {Renschler}, {Renzi},
  {Resconi}, {Rhode}, {Richman}, {Robertson}, {Rongen}, {Rott}, {Ruhe},
  {Ryckbosch}, {Rysewyk}, {Safa}, {Sanchez Herrera}, {Sandrock}, {Sand roos},
  {Santander}, {Sarkar}, {Sarkar}, {Satalecka}, {Schaufel}, {Schieler},
  {Schlunder}, {Schmidt}, {Schneider}, {Schneider}, {Schr{\"o}der},
  {Schumacher}, {Sclafani}, {Seckel}, {Seunarine}, {Shefali}, {Silva},
  {Snihur}, {Soedingrekso}, {Soldin}, {Song}, {Spiczak}, {Spiering},
  {Stachurska}, {Stamatikos}, {Stanev}, {Stein}, {Stettner}, {Steuer},
  {Stezelberger}, {Stokstad}, {St{\"o}{\ss}l}, {Strotjohann}, {St{\"u}rwald},
  {Stuttard}, {Sullivan}, {Taboada}, {Tenholt}, {Ter-Antonyan}, {Terliuk},
  {Tilav}, {Tollefson}, {Tomankova}, {T{\"o}nnis}, {Toscano}, {Tosi},
  {Trettin}, {Tselengidou}, {Tung}, {Turcati}, {Turcotte}, {Turley}, {Ty},
  {Unger}, {Unland Elorrieta}, {Usner}, {Vandenbroucke}, {Van Driessche}, {van
  Eijk}, {van Eijndhoven}, {van Santen}, {Verpoest}, {Vraeghe}, {Walck},
  {Wallace}, {Wallraff}, {Wandkowsky}, {Watson}, {Weaver}, {Weindl}, {Weiss},
  {Weldert}, {Wendt}, {Werthebach}, {Whelan}, {Whitehorn}, {Wiebe}, {Wiebusch},
  {Wille}, {Williams}, {Wills}, {Wolf}, {Wood}, {Wood}, {Woschnagg}, {Wrede},
  {Xu}, {Xu}, {Xu}, {Yanez}, {Yodh}, {Yoshida}, {Yuan}, \&
  {Z{\"o}cklein}}]{2020arXiv200109520I}
{IceCube Collaboration}, {Aartsen}, M.~G., {Ackermann}, M., {et~al.} 2020,
  arXiv e-prints, arXiv:2001.09520.
\newblock \doarXiv{2001.09520}

\bibitem[{{Keivani} {et~al.}(2018){Keivani}, {Murase}, {Petropoulou}, {Fox},
  {Cenko}, {Chaty}, {Coleiro}, {DeLaunay}, {Dimitrakoudis}, {Evans}, {Kennea},
  {Marshall}, {Mastichiadis}, {Osborne}, {Santand er}, {Tohuvavohu}, \&
  {Turley}}]{Keivani2018}
{Keivani}, A., {Murase}, K., {Petropoulou}, M., {et~al.} 2018, \apj, 864, 84,
  \dodoi{10.3847/1538-4357/aad59a}

\bibitem[{{Kirk} \& {Mastichiadis}(1992)}]{1992Natur.360..135K}
{Kirk}, J.~G., \& {Mastichiadis}, A. 1992, \nat, 360, 135,
  \dodoi{10.1038/360135a0}

\bibitem[{{Krauss} {et~al.}(2020){Krauss}, {Gregoire}, {Fox}, {Kennea}, \&
  {Evans}}]{2020Atel13395....1K}
{Krauss}, F., {Gregoire}, T., {Fox}, D.~B., {Kennea}, J., \& {Evans}, P. 2020,
  The Astronomer's Telegram, 13395, 1

\bibitem[{{Krimm} {et~al.}(2013){Krimm}, {Holland}, {Corbet}, {Pearlman},
  {Romano}, {Kennea}, {Bloom}, {Barthelmy}, {Baumgartner}, {Cummings},
  {Gehrels}, {Lien}, {Markwardt}, {Palmer}, {Sakamoto}, {Stamatikos}, \&
  {Ukwatta}}]{2013ApJS..209...14K}
{Krimm}, H.~A., {Holland}, S.~T., {Corbet}, R.~H.~D., {et~al.} 2013, \apjs,
  209, 14, \dodoi{10.1088/0067-0049/209/1/14}

\bibitem[{{Liodakis} {et~al.}(2019){Liodakis}, {Peirson}, \&
  {Romani}}]{2019ApJ...880...29L}
{Liodakis}, I., {Peirson}, A.~L., \& {Romani}, R.~W. 2019, \apj, 880, 29,
  \dodoi{10.3847/1538-4357/ab2719}

\bibitem[{{Mannheim}(1993)}]{1993A&A...269...67M}
{Mannheim}, K. 1993, \aap, 269, 67.
\newblock \doarXiv{astro-ph/9302006}

\bibitem[{{Mannheim} \& {Biermann}(1992)}]{1992A&A...253L..21M}
{Mannheim}, K., \& {Biermann}, P.~L. 1992, \aap, 253, L21

\bibitem[{{Mannheim} {et~al.}(1991){Mannheim}, {Biermann}, \&
  {Kruells}}]{1991A&A...251..723M}
{Mannheim}, K., {Biermann}, P.~L., \& {Kruells}, W.~M. 1991, \aap, 251, 723

\bibitem[{{Mannheim} \& {Schlickeiser}(1994)}]{1994A&A...286..983M}
{Mannheim}, K., \& {Schlickeiser}, R. 1994, \aap, 286, 983.
\newblock \doarXiv{astro-ph/9402042}

\bibitem[{{Mastichiadis} {et~al.}(2020){Mastichiadis}, {Florou}, {Kefala},
  {Boula}, \& {Petropoulou}}]{2020MNRAS.495.2458M}
{Mastichiadis}, A., {Florou}, I., {Kefala}, E., {Boula}, S.~S., \&
  {Petropoulou}, M. 2020, \mnras, 495, 2458, \dodoi{10.1093/mnras/staa1308}

\bibitem[{{Mastichiadis} \& {Kirk}(1995)}]{1995A&A...295..613M}
{Mastichiadis}, A., \& {Kirk}, J.~G. 1995, \aap, 295, 613

\bibitem[{{Mastichiadis} {et~al.}(2013){Mastichiadis}, {Petropoulou}, \&
  {Dimitrakoudis}}]{2013MNRAS.434.2684M}
{Mastichiadis}, A., {Petropoulou}, M., \& {Dimitrakoudis}, S. 2013, \mnras,
  434, 2684, \dodoi{10.1093/mnras/stt1210}

\bibitem[{{McEnery} {et~al.}(2019){McEnery}, {van der Horst}, {Dominguez},
  {Moiseev}, {Marcowith}, {Harding}, {Lien}, {Giuliani}, {Inglis}, {Ansoldi},
  {Stamerra}, {Manousakis}, {Strong}, {Bambi}, {Patricelli}, {Baring},
  {Barrio}, {Bastieri}, {Fields}, {Beacom}, {Beckmann}, {Bednarek}, {Rani},
  {Boggs}, {Bolotnikov}, {Cenko}, {Buckley}, {Grefenstette}, {Hui}, {Pittori},
  {Prescod-Weinstein}, {Shrader}, {Gouiffes}, {Kierans}, {Wilson-Hodge},
  {D'Ammando}, {Castro}, {Kocveski}, {Gasparrini}, {Thompson}, {Williams}, {De
  Angelis}, {Bernard}, {Digel}, {Morcuende}, {Charles}, {Bissaldi}, {Hays},
  {Ferrara}, {Bozzo}, {Grove}, {Wulf}, {Bottacini}, {Caroli}, {Kislat},
  {Oikonomou}, {Giordano}, {Longo}, {Fryer}, {Fukazawa}, {Georganopoulos}, {De
  Nolfo}, {Vianello}, {Kanbach}, {Younes}, {Blumer}, {Hartmann}, {Hernanz},
  {Takahashi}, {Li}, {Agudo}, {Moskalenko}, {Stumke}, {Grenier}, {Smith},
  {Rodi}, {Perkins}, {Gelfand}, {Holder}, {Knodlseder}, {Kopp}, {Lenain},
  {{\'A}lvarez}, {Metcalfe}, {Krizmanic}, {Stephen}, {Hewitt}, {Mitchell},
  {Harding}, {Tomsick}, {Racusin}, {Finke}, {Kargaltsev}, {Klimenko},
  {Krawczynski}, {Smith}, {Kubo}, {Di Venere}, {Marcotulli}, {Lommler},
  {Parker}, {Baldini}, {Foffano}, {Zampieri}, {Tibaldo}, {Petropoulou},
  {Ajello}, {Meyer}, {L{\'o}pez}, {McConnell}, {Boettcher}, {Cardillo},
  {Martinez}, {Kerr}, {Mazziotta}, {McEnery}, {Di Mauro}, {Wood}, {Meyer},
  {Briggs}, {De Becker}, {Lovellette}, {Doro}, {Sanchez-Conde}, {Moss},
  {Mizuno}, {Rib{\'o}}, {Nakazawa}, {Neilson}, {Auricchio}, {Omodei},
  {Oberlack}, {Ohno}, {Orland o}, {Otte}, {Coppi}, {Bloser}, {Zhang},
  {Laurent}, {Pohl}, {Prand ini}, {Shawhan}, {Caputo}, {Campana}, {Rando},
  {Woolf}, {Johnson}, {Mignani}, {Walter}, {Ojha}, {da Silva}, {Dietrich},
  {Funk}, {Zane}, {Anton}, {Buson}, {Cutini}, {Saz Parkinson}, {Schirato},
  {Griffin}, {Kaufmann}, {Stawarz}, {Ciprini}, {Del Sordo}, {Jones}, {Guiriec},
  {Tajima}, {Cheung}, {The}, {Venters}, {Porter}, {Linden}, {Barres}, {Paliya},
  {Bozhilov}, {Vestrand}, {Tatischeff}, {Chen}, {Wang}, {Tanaka}, {Uhm},
  {Zhang}, {Zimmer}, {Zoglauer}, \& {Wadiasingh}}]{2019BAAS...51g.245M}
{McEnery}, J., {van der Horst}, A., {Dominguez}, A., {et~al.} 2019, in \baas,
  Vol.~51, 245.
\newblock \doarXiv{1907.07558}

\bibitem[{{M{\"u}cke} \& {Protheroe}(2001)}]{2001APh....15..121M}
{M{\"u}cke}, A., \& {Protheroe}, R.~J. 2001, Astroparticle Physics, 15, 121,
  \dodoi{10.1016/S0927-6505(00)00141-9}

\bibitem[{{M{\"u}cke} {et~al.}(2003){M{\"u}cke}, {Protheroe}, {Engel},
  {Rachen}, \& {Stanev}}]{2003APh....18..593M}
{M{\"u}cke}, A., {Protheroe}, R.~J., {Engel}, R., {Rachen}, J.~P., \& {Stanev},
  T. 2003, Astroparticle Physics, 18, 593,
  \dodoi{10.1016/S0927-6505(02)00185-8}

\bibitem[{{Murase} {et~al.}(2016){Murase}, {Guetta}, \&
  {Ahlers}}]{2016PhRvL.116g1101M}
{Murase}, K., {Guetta}, D., \& {Ahlers}, M. 2016, \prl, 116, 071101,
  \dodoi{10.1103/PhysRevLett.116.071101}

\bibitem[{{Murase} {et~al.}(2018){Murase}, {Oikonomou}, \&
  {Petropoulou}}]{2018ApJ...865..124M}
{Murase}, K., {Oikonomou}, F., \& {Petropoulou}, M. 2018, \apj, 865, 124,
  \dodoi{10.3847/1538-4357/aada00}

\bibitem[{{O'Dell} {et~al.}(2018){O'Dell}, {Baldini}, {Bellazzini}, {Costa},
  {Elsner}, {Kaspi}, {Kolodziejczak}, {Latronico}, {Marshall}, {Matt},
  {Mulieri}, {Ramsey}, {Romani}, {Soffitta}, {Tennant}, {Weisskopf}, {Allen},
  {Amici}, {Antoniak}, {Attina}, {Bachetti}, {Barbanera}, {Baumgartner},
  {Bladt}, {Bongiorno}, {Borotto}, {Brooks}, {Bussinger}, {Bygott},
  {Cavazzuti}, {Ceccanti}, {Citraro}, {Deininger}, {Del Monte}, {Dietz}, {Di
  Lalla}, {Di Persio}, {Donnarumma}, {Erickson}, {Evangelista}, {Fabiani},
  {Ferrazzoli}, {Foster}, {Giusti}, {Gunji}, {Guy}, {Johnson}, {Kalinowski},
  {Kelley}, {Kilaru}, {Lefevre}, {Maldera}, {Manfreda}, {Marengo},
  {Masciarelli}, {McEachen}, {Mereu}, {Minuti}, {Mitchell}, {Mitchell},
  {Mitsuishi}, {Morbidini}, {Mosti}, {Nasimi}, {Negri}, {Orsini}, {Osborne},
  {Pavelitz}, {Pentz}, {Perri}, {Pesce-Rollins}, {Peterson}, {Piazzolla},
  {Pieraccini}, {Pilia}, {Pinchera}, {Puccetti}, {Ranganathan}, {Read},
  {Rubini}, {Santoli}, {Sarra}, {Schindhelm}, {Sciortino}, {Seckar},
  {Sgr{\`o}}, {Smith}, {Speegle}, {Tamagawa}, {Tardiola}, {Tobia}, {Tortosa},
  {Trois}, {Weddendorf}, {Wedmore}, \& {Zanetti}}]{IXPE2018}
{O'Dell}, S.~L., {Baldini}, L., {Bellazzini}, R., {et~al.} 2018, in Society of
  Photo-Optical Instrumentation Engineers (SPIE) Conference Series, Vol. 10699,
  Space Telescopes and Instrumentation 2018: Ultraviolet to Gamma Ray, 106991X,
  \dodoi{10.1117/12.2314146}

\bibitem[{{Paiano} {et~al.}(2020){Paiano}, {Falomo}, {Padovani}, {Giommi},
  {Gargiulo}, {Uslenghi}, {Rossi}, \& {Treves}}]{2020MNRAS.495L.108P}
{Paiano}, S., {Falomo}, R., {Padovani}, P., {et~al.} 2020, \mnras, 495, L108,
  \dodoi{10.1093/mnrasl/slaa056}

\bibitem[{{Paliya} {et~al.}(2018){Paliya}, {Zhang}, {B{\"o}ttcher}, {Ajello},
  {Dom{\'\i}nguez}, {Joshi}, {Hartmann}, \& {Stalin}}]{2018ApJ...863...98P}
{Paliya}, V.~S., {Zhang}, H., {B{\"o}ttcher}, M., {et~al.} 2018, \apj, 863, 98,
  \dodoi{10.3847/1538-4357/aad1f0}

\bibitem[{{Petropoulou} {et~al.}(2014){Petropoulou}, {Dimitrakoudis},
  {Mastichiadis}, \& {Giannios}}]{2014MNRAS.444.2186P}
{Petropoulou}, M., {Dimitrakoudis}, S., {Mastichiadis}, A., \& {Giannios}, D.
  2014, \mnras, 444, 2186, \dodoi{10.1093/mnras/stu1362}

\bibitem[{{Petropoulou} {et~al.}(2015){Petropoulou}, {Dimitrakoudis},
  {Padovani}, {Mastichiadis}, \& {Resconi}}]{Petropoulou2015}
{Petropoulou}, M., {Dimitrakoudis}, S., {Padovani}, P., {Mastichiadis}, A., \&
  {Resconi}, E. 2015, \mnras, 448, 2412, \dodoi{10.1093/mnras/stv179}

\bibitem[{{Petropoulou} \& {Mastichiadis}(2015)}]{2015MNRAS.447...36P}
{Petropoulou}, M., \& {Mastichiadis}, A. 2015, \mnras, 447, 36,
  \dodoi{10.1093/mnras/stu2364}

\bibitem[{{Petropoulou} {et~al.}(2020{\natexlab{a}}){Petropoulou}, {Oikonomou},
  {Mastichiadis}, {Murase}, {Padovani}, {Vasilopoulos}, \&
  {Giommi}}]{2020ApJ...899..113P}
{Petropoulou}, M., {Oikonomou}, F., {Mastichiadis}, A., {et~al.}
  2020{\natexlab{a}}, \apj, 899, 113, \dodoi{10.3847/1538-4357/aba8a0}

\bibitem[{{Petropoulou} {et~al.}(2020{\natexlab{b}}){Petropoulou}, {Murase},
  {Santander}, {Buson}, {Tohuvavohu}, {Kawamuro}, {Vasilopoulos}, {Negoro},
  {Ueda}, {Siegel}, {Keivani}, {Kawai}, {Mastichiadis}, \&
  {Dimitrakoudis}}]{2020ApJ...891..115P}
{Petropoulou}, M., {Murase}, K., {Santander}, M., {et~al.} 2020{\natexlab{b}},
  \apj, 891, 115, \dodoi{10.3847/1538-4357/ab76d0}

\bibitem[{{Ray} {et~al.}(2019){Ray}, {Arzoumanian}, {Ballantyne}, {Bozzo},
  {Brandt}, {Brenneman}, {Chakrabarty}, {Christophersen}, {DeRosa}, {Feroci},
  {Gendreau}, {Goldstein}, {Hartmann}, {Hernanz}, {Jenke}, {Kara}, {Maccarone},
  {McDonald}, {Nowak}, {Phlips}, {Remillard}, {Stevens}, {Tomsick}, {Watts},
  {Wilson-Hodge}, {Wood}, {Zane}, {Ajello}, {Alston}, {Altamirano}, {Antoniou},
  {Arur}, {Ashton}, {Auchettl}, {Ayres}, {Bachetti}, {Balokovic}, {Baring},
  {Baykal}, {Begelman}, {Bhat}, {Bogdanov}, {Briggs}, {Bulbul}, {Bult},
  {Burns}, {Cackett}, {Campana}, {Caspi}, {Cavecchi}, {Chenevez}, {Cherry},
  {Corbet}, {Corcoran}, {Corsi}, {Degenaar}, {Drake}, {Eikenberry}, {Enoto},
  {Fragile}, {Fuerst}, {Gandhi}, {Garcia}, {Goldstein}, {Gonzalez},
  {Grefenstette}, {Grinberg}, {Grossan}, {Guillot}, {Guver}, {Haggard},
  {Heinke}, {Heinz}, {Hemphill}, {Homan}, {Hui}, {Huppenkothen}, {Ingram},
  {Irwin}, {Jaisawal}, {Jaodand}, {Kalemci}, {Kaplan}, {Keek}, {Kennea},
  {Kerr}, {van der Klis}, {Kocevski}, {Koss}, {Kowalski}, {Lai}, {Lamb},
  {Laycock}, {Lazio}, {Lazzati}, {Longcope}, {Loewenstein}, {Maitra}, {Majid},
  {Maksym}, {Malacaria}, {Margutti}, {Martindale}, {McHardy}, {Meyer},
  {Middleton}, {Miller}, {Miller}, {Motta}, {Neilsen}, {Nelson}, {Noble},
  {O'Brien}, {Osborne}, {Osten}, {Ozel}, {Palliyaguru}, {Pasham}, {Patruno},
  {Pelassa}, {Petropoulou}, {Pilia}, {Pohl}, {Pooley}, {Prescod-Weinstein},
  {Psaltis}, {Raaijmakers}, {Reynolds}, {Riley}, {Salvesen}, {Santangelo},
  {Scaringi}, {Schanne}, {Schnittman}, {Smith}, {Smith}, {Snios}, {Steiner},
  {Steiner}, {Stella}, {Strohmayer}, {Sun}, {Tauris}, {Taylor}, {Tohuvavohu},
  {Vacchi}, {Vasilopoulos}, {Veledina}, {Walsh}, {Weinberg}, {Wilkins},
  {Willingale}, {Wilms}, {Winter}, {Wolff}, {in 't Zand}, {Zezas}, {Zhang}, \&
  {Zoghbi}}]{2019arXiv190303035R}
{Ray}, P.~S., {Arzoumanian}, Z., {Ballantyne}, D., {et~al.} 2019, arXiv
  e-prints, arXiv:1903.03035.
\newblock \doarXiv{1903.03035}

\bibitem[{{Reimer} {et~al.}(2019){Reimer}, {B{\"o}ttcher}, \&
  {Buson}}]{2019ApJ...881...46R}
{Reimer}, A., {B{\"o}ttcher}, M., \& {Buson}, S. 2019, \apj, 881, 46,
  \dodoi{10.3847/1538-4357/ab2bff}

\bibitem[{{Rodrigues} {et~al.}(2019){Rodrigues}, {Gao}, {Fedynitch},
  {Palladino}, \& {Winter}}]{2019ApJ...874L..29R}
{Rodrigues}, X., {Gao}, S., {Fedynitch}, A., {Palladino}, A., \& {Winter}, W.
  2019, \apjl, 874, L29, \dodoi{10.3847/2041-8213/ab1267}

\bibitem[{{Rodrigues} {et~al.}(2020){Rodrigues}, {Garrappa}, {Gao}, {Paliya},
  {Franckowiak}, \& {Winter}}]{2020arXiv200904026R}
{Rodrigues}, X., {Garrappa}, S., {Gao}, S., {et~al.} 2020, arXiv e-prints,
  arXiv:2009.04026.
\newblock \doarXiv{2009.04026}

\bibitem[{{Stecker} {et~al.}(1991){Stecker}, {Done}, {Salamon}, \&
  {Sommers}}]{1991PhRvL..66.2697S}
{Stecker}, F.~W., {Done}, C., {Salamon}, M.~H., \& {Sommers}, P. 1991, \prl,
  66, 2697, \dodoi{10.1103/PhysRevLett.66.2697}

\bibitem[{{Zhang} {et~al.}(2020){Zhang}, {Petropoulou}, {Murase}, \&
  {Oikonomou}}]{2020ApJ...889..118Z}
{Zhang}, B.~T., {Petropoulou}, M., {Murase}, K., \& {Oikonomou}, F. 2020, \apj,
  889, 118, \dodoi{10.3847/1538-4357/ab659a}

\bibitem[{{Zhang} \& {B{\"o}ttcher}(2013)}]{2013ApJ...774...18Z}
{Zhang}, H., \& {B{\"o}ttcher}, M. 2013, \apj, 774, 18,
  \dodoi{10.1088/0004-637X/774/1/18}

\bibitem[{{Zhang} {et~al.}(2019){Zhang}, {Fang}, {Li}, {Giannios},
  {B{\"o}ttcher}, \& {Buson}}]{2019ApJ...876..109Z}
{Zhang}, H., {Fang}, K., {Li}, H., {et~al.} 2019, \apj, 876, 109,
  \dodoi{10.3847/1538-4357/ab158d}

\end{thebibliography}
\bibliographystyle{aasjournal}



\end{document}